\RequirePackage[switch,columnwise]{lineno}
\documentclass[prb,amsmath,groupedaddress,floatfix,twocolumn]{revtex4-1}

\usepackage{graphicx}
\usepackage{amssymb}
\usepackage{longtable}
\usepackage{rotating}
\usepackage{array}
\usepackage{chemist}
\usepackage{multirow}
\usepackage{makecell}
\usepackage{booktabs}
\usepackage{xcolor}

\begin{document}


\title{Energetics of oxygen-octahedra rotations in perovskite oxides
  from first principles}

\author{Peng Chen,$^{1,2}$ Mathieu N. Grisolia,$^{3}$ Hong Jian
  Zhao,$^{1}$ Otto E. Gonz\'alez-V\'azquez,$^{4,5}$
  L. Bellaiche,$^{6}$ Manuel Bibes,$^{3}$ Bang-Gui Liu,$^{2}$ and
  Jorge \'I\~niguez$^{1,5}$}

\address{
  $^{1}$Materials Research and Technology Department, Luxembourg
  Institute of Science and Technology (LIST), 5 avenue des
  Hauts-Fourneaux, L-4362 Esch/Alzette, Luxembourg \\
$^{2}$Beijing National Laboratory for Condensed Matter Physics,
  Institute of Physics Chinese Academy of Science, Beijing 100190,
  China\\
$^{3}$Unit\'e Mixte de Physique, CNRS, Thales, Universit\'e Paris Sud,
  Universit\'e Paris-Saclay, 1 avenue A. Fresnel, 91767, Palaiseau,
  France\\
$^{4}$Scientific Computing \& Software for Experiments Department,
  Sincrotrone Elettra, 34149 Basovizza, Trieste, Italy\\
$^{5}$Institut de Ci\`encia de Materials de Barcelona (ICMAB-CSIC),
  Campus UAB, 08193 Bellaterra, Spain\\
  $^{6}$Physics Department and Institute for Nanoscience and
  Engineering, University of Arkansas, Fayetteville, Arkansas 72701,
  USA
}


\begin{abstract}
We use first-principles methods to investigate the energetics of
oxygen-octahedra rotations in {\sl AB}O$_{3}$ perovskite oxides. We
focus on the short-period, perfectly antiphase or in-phase, tilt
patterns that characterize the structure of most compounds and control
their physical (e.g., conductive, magnetic) properties. Based on an
analytical form of the relevant potential energy surface, we discuss
the conditions for the stability of various polymorphs presenting
different rotation patterns, and obtain numerical results for a
collection of thirty-five representative materials. Our results reveal
the mechanisms responsible for the frequent occurrence of a particular
structure that combines antiphase and in-phase rotations, i.e., the
orthorhombic $Pbnm$ phase displayed by about half of all perovskite
oxides, as well as by many non-oxidic perovskites. In essence, the
$Pbnm$ phase benefits from the simultaneous occurrence of antiphase
and in-phase tilt patterns that compete with each other, but not as
strongly as to be mutually exclusive. We also find that secondary
antipolar modes, involving the {\sl A} cations, contribute to weaken
the competition between tilts of different types, and thus play a key
role in the stabilization of the $Pbnm$ structure. Our results thus
confirm and better explain previous observations for particular
compounds in the literature. Interestingly, we also find that strain
effects, which are known to be a major factor governing phase
competition in related (e.g., ferroelectric) perovskite oxides, play
no essential role as regards the relative stability of different
rotational polymorphs. Further, we discuss why the $Pbnm$ structure
stops being the ground state in two opposite limits -- namely, for
large and small {\sl A} cations --, showing that very different
effects become relevant in each case. Our work thus provides a
comprehensive discussion and reference data on these all-important and
abundant materials, which will be useful to better understand existing
compounds as well as to identify new strategies for materials
engineering.
\end{abstract}


\maketitle




\section{Introduction}

\begin{figure}
\includegraphics[width=1.00\linewidth]{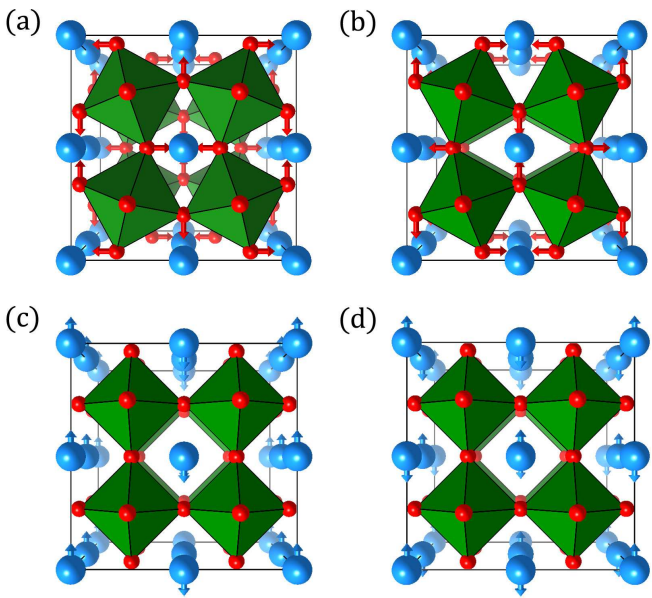}
  \caption{Sketch of the perovskite structure and the most important
    distortion modes discussed in this work. Blue and red circles
    represent {\sl A} and oxygen atoms, respectively, while we show
    the O$_{6}$ octahedra centered on the {\sl B} atoms in green. (a)
    Pattern of antiphase O$_{6}$ rotations, about the $z$ pseudo-cubic
    axis (perpendicular to the page). (b) In-phase rotations about the
    $z$ axis. (c) Antipolar {\sl A}-cation displacements modulated
    according to the ${\bf q}_{X} = \pi/a(1,0,0)$ wave vector
    (horizontal direction in the figure). Here, $a$ is the lattice
    constant of the 5-atom elemental perovskite cell for the reference
    cubic phase. We show the ${\bf q}_{X}$ modulation in the figure
    for clarity; yet, the $a^{-}a^{-}c^{+}$ polymorph discussed in the
    text displays an equivalent antipolar distortion modulated by
    ${\bf q}_{Z} = \pi/a(0,0,1)$. (d) Antipolar {\sl A}-cation
    displacements modulated according to the ${\bf q}_{R} =
    \pi/a(1,1,1)$ wave vector.}
\label{fig:sketch-dists}
\end{figure}

Most {\sl AB}O$_{3}$ perovskite oxides present structures that are
distorted versions of the ideal cubic phase. In the vast majority of
compounds, this distortion is characterized by concerted, short-period
rotations of the O$_{6}$ oxygen octahedra that constitute the basic
building block of the perovskite lattice.\cite{glazer72,lufaso01} The
most common rotation patterns can be described as being exactly
antiphase [usually denoted with a ``$-$'' sign, see
  Fig.~\ref{fig:sketch-dists}(a)] or in-phase [``$+$'', see
  Fig.~\ref{fig:sketch-dists}(b)], and often appear together. Indeed,
about half of the perovskite oxides present the so-called
GdFeO$_{3}$-type structure,\cite{lufaso01} an orthorhombic polymorph
with $Pbnm$ space group characterized by antiphase rotations about the
[110] pseudo-cubic axis and in-phase rotations about [001]. This
structure is usually termed ``$a^{-}a^{-}c^{+}$'' in the notation
introduced by Glazer,\cite{glazer72} which is self-explanatory. Other
common structures present only antiphase tilts, and typically adopt
tetragonal ($a^{-}b^{0}b^{0}$, $I4/mcm$ space group) and rhombohedral
($a^{-}a^{-}a^{-}$, $R\bar{3}c$ space group)
symmetries.\cite{glazer72} All these purely-rotational phases are
sometimes called {\em antiferrodistortive} (AFD), and all are
ferroelastic.\cite{salje-book1993,salje12} The O$_{6}$ rotations
sometimes coexist with other primary distortions, as e.g. cation
off-centerings that give rise to
ferroelectricity.\cite{lines-book1977,rabe-book2007} Notably, this is
the case of materials like room-temperature multiferroic
BiFeO$_{3}$.\cite{catalan09} Such a coexistence is rare, though, as
the ferroelectric (FE) and AFD instabilities are known to compete
against each other in the most typical
situations.\cite{zhong95b,kornev06,wojdel13,gu17} Hence, most FE
perovskites (e.g., BaTiO$_{3}$, PbTiO$_{3}$, KNbO$_{3}$) do not
present any O$_{6}$ tilts at all.

The tendency of perovskites to display O$_{6}$ rotations is usually
explained in terms of the so-called tolerance
factor\cite{goldschmidt26}
\begin{equation}
  t = \frac{R_{A} + R_{\rm O}}{\sqrt{2}(R_{B} + R_{\rm O})} \; ,
\end{equation}
where $R_{A}$, $R_{B}$, and $R_{\rm O}$ are the nominal ionic radii of
the {\sl A}, {\sl B}, and O species, respectively. (Which we typically
take from Ref.~\onlinecite{shannon76}.) This quantity is defined so
that $t = 1$ corresponds to the ideal case in which rigid spheres with
the radii of the corresponding ions are perfectly stuck in the cubic
perovskite lattice. In contrast, if $t\neq 1$, the cubic lattice is in
principle unstable. In particular, $t < 1$ values imply that the {\sl
  A} cation is small as compared with the cage of surrounding oxygens,
so that, most likely, a distortion will occur to optimize the {\sl
  A}--O bond distances. Typically, rigid rotations of the O$_{6}$
octahedra are the most favorable possible distortions, and thus
structures with tilts abound.

Octahedral tilts characterize many of the most important perovskite
compounds, including all the manganites\cite{dagotto-book2003} and
nickelates\cite{catalan08,middey16} that attract great interest
because of their peculiar magnetic, conductive, and magnetoresistive
properties. Most of the today much-studied iridates,\cite{rau16} where
Ir is a relative large cation at the {\sl B} site of the perovskite
lattice, present tilted phases as well, and so do the
orthoferrites\cite{bousquet16,zhao16,zhao17} that have recently gained
renewed attention because of their multiferroic and spin-dynamical
properties. Moving beyond the oxides, there are plenty of materials
families displaying tilted phases, as e.g. the novel hybrid
perovskites with incredible photovoltaic properties.\cite{gratzel14}
Interestingly, the tilting distortion is known to be the key
structural factor controlling the electronic properties of all these
compounds, as it determines the overlap between the orbitals of the
anion and {\sl B}-site cation (often a transition metal in perovskite
oxides).\cite{goodenough-book1963} Hence, today there is a great
interest in understanding the details of such distortions, and in
exploring new possibilities to tune them, as illustrated by many
recent works on epitaxial oxide thin films.\cite{schlom07,schlom14}

Additionally, it has been recently demonstrated that tilted structures
provide an unconventional, and in some respects advantageous, strategy
to obtain polar, potentially ferroelectric, materials. This so-called
{\em hybrid improper ferroelectricity}\cite{benedek11,bousquet08} has
been predicted in short-period superlattices based on $Pbnm$
compounds,\cite{rondinelli12,mulder13,zhao14c} and could be a
convenient route to obtain elusive effects such as room-temperature
magnetoelectricity.\cite{benedek12,zanolli13,zhao14a} Experimental
demonstrations of this exotic form of ferroelectricity are starting to
appear,\cite{oh15,kim16} and highlight once again the importance of
understanding O$_{6}$ rotational patterns in perovskites, even in
contexts where their relevance was difficult to anticipate {\sl a
  priori}.

Given the obvious interest of these tilting distortions, it is
surprising to note that they remain relatively poorly studied,
especially when compared with the FE instabilities of compounds like
BaTiO$_{3}$ or PbTiO$_{3}$. For example, while there is an exhaustive
crystallographic literature on O$_{6}$-rotational
patterns,\cite{glazer72,glazer75,woodward97a,woodward97b,stokes02}
there are very few phenomenological works discussing the energetics
and phase transitions of materials with tilted phases. Historically,
this is probably related to the fact that these structures (especially
those with the $Pbnm$ space group) tend to be very stable in wide
ranges of temperature and pressure, including ambient conditions,
which renders them a relatively uninteresting subject of study {\sl a
  priori}. First-principles theory is somewhat underdeveloped as
well. Admittedly, there are a number of recent works on how to control
O$_{6}$ rotations by epitaxial strain in thin films of specific
compounds,\cite{rondinelli12} and tilts are the focus of other
investigations in various contexts. Yet, in our view we are still
missing a thorough first-principles study of these instabilities, and
of why some rotational polymorphs prevail over others. For the sake of
comparison, in the case of ferroelectric perovskites, the basic
first-principles works at the origin of our current understanding,
which enabled much of the later progress in FE thin films and strain
engineering, were laid out in the early
90's.\cite{cohen92,kingsmith94,posternak94} In our view, especially
relevant were seminal contributions as that of King-Smith and
Vanderbilt in Ref.~\onlinecite{kingsmith94}; these authors ran a
comparative study of a group of representative compounds, and
quantified trends in the framework of a simple energy model, which
allowed them to rationalize the factors controlling why apparently
similar materials present different ferroelectric phases. Our purpose
in this work is to provide the same kind of description and insights
in what regards octahedral tilts in perovskite oxides.

The paper is organized as follows. In Section~\ref{sec:formalism} we
introduce the formalism that allows us to model the potential energy
surface (PES) of a perovskite, around the reference cubic structure,
as a function of antiphase and in-phase O$_{6}$ rotations and cell
strains. We discuss the relevant critical points of the PES and their
stability. In Section~\ref{sec:methods} we describe our
first-principles computational approach, and justify the choice of the
materials considered in this investigation. In
Section~\ref{sec:results} we present and discuss our computational
results. Finally, in Section~\ref{sec:conclusions} we summarize our
conclusions.

\section{Formalism}
\label{sec:formalism}

 In this Section we introduce a general model to describe
  the PES of any perovskite, around the ideal cubic phase, as a
  function of short-period rotations of the oxygen octahedra and
  macroscopic strains. This approach is a direct application to
  rotational distortions of the methodology described in
  Ref.~\onlinecite{kingsmith94}, and our derivations are essentially
  identical to those described in Ref.~\onlinecite{gu12} within an
  investigation of CaTiO$_{3}$.

The formalism below applies to the idealized case of an
  infinite, periodic crystal that is free of defects. Further, some
  important physical effects are not treated in our theory. For
  example, we ignore the possibility of having spin-ordering
  transitions -- as occurring, e.g., in the considered orthoferrites
  and orthochromites\cite{bousquet16,white69} -- and the way those
  could affect the energetics of the tilting distorions; in fact, we
  implicitly assume that the materials always remain in their magnetic
  ground state. Thus, while these simplifications are acceptable for
  the present study, one should bear in mind that, in order to address
  more complex phenomena, the present models would need to be
  extended. (See Refs.~\onlinecite{zhao16,zhao17} for examples of
  models including magnetostructural couplings.)

\subsection{Relevant potential energy surface}

We express the energy as a Taylor series, in terms of the relevant
structural distortions, around a reference cubic structure. More
precisely, we write:
\begin{equation}
\begin{split}
  E = & E_{0} + E_{\rm s}(\{\eta_{a}\}) + E_{r}(\{r_{\alpha}\}) +
  E_{m}(\{m_{\beta}\}) \\ & + E_{\rm
    int}(\{r_{\alpha}\},\{m_{\beta}\}) + E_{\rm
    sp}(\{\eta_{a}\},\{r_{\alpha}\},\{m_{\beta}\}) \, ,
\end{split}
\label{eq:energy}
\end{equation}
where $E_{0}$ is the energy of the ideal perovskite cubic phase with a
5-atom periodically-repeated cell, as obtained from a
symmetry-constrained first-principles structural relaxation; $E_{\rm
  s}$ is the elastic energy as a function of the homogeneous strains
$\eta_{a}$, with $a = 1, ..., 6$ in Voigt notation;\cite{nye-book1985}
$E_{r}$ is the energy associated to antiphase rotations of the oxygen
octahedra about the $\alpha = x, y, z$ pseudo-cubic axes, as
quantified by $r_{\alpha}$; $E_{m}$ is the analogous function for the
in-phase O$_{6}$ rotations, as given by $m_{\beta}$ with $\beta =
x,y,z$; $E_{\rm int}$ gathers the anharmonic interactions between
antiphase and in-phase tilts; finally, $E_{\rm sp}$ -- where ``sp''
stands for strain-phonon -- accounts for the coupling between AFD modes
and strains. Let us note that the rotation amplitudes $r_{\alpha}$ and
$m_{\beta}$ are associated to distortion patterns as those indicated
in Figs.~\ref{fig:sketch-dists}(a) and \ref{fig:sketch-dists}(b),
respectively. Note also that our choice of notation for the antiphase
($r_{\alpha}$) and in-phase ($m_{\beta}$) rotations reflects the fact
that these distortions are respectively associated with the $R$ [${\bf
    q}_{R} = \pi/a (1,1,1)$] and $M$ [${\bf q}_{M} = \pi/a (1,1,0)$
  for in-phase rotations about the $z$ axis] $q$-points of the
Brillouin zone corresponding to the ideal 5-atom perovskite cell. Note
that ${\bf q}_{R}$ and ${\bf q}_{M}$ are zone-boundary wave vectors,
and $a$ is the lattice constant of the 5-atom elemental cell as
obtained from a symmetry-constrained relaxation of the cubic reference
structure.

This energy must be invariant with respect to the symmetry operations
of the cubic phase, which greatly simplifies its form. In the
following we write the various terms, truncating the Taylor series at
the lowest order that makes it possible to discuss the structural
instabilities and their most relevant couplings. Thus, for example,
for the elastic energy we have
\begin{equation}
\begin{split}
E_{\rm s} = & \frac{1}{2}
C_{11}(\eta_{1}^{2}+\eta_{2}^{2}+\eta_{3}^{2})\\
& +C_{12}(\eta_{1}\eta_{2}+\eta_{2}\eta_{3}+\eta_{3}\eta_{1})\\
&+\frac{1}{2}C_{44}(\eta_{4}^{2}+\eta_{5}^{2}+\eta_{6}^{2}) \, ,
\end{split}
\end{equation}
where the $C_{ab}$ parameters are the usual elastic constants. Note
that, because of the cubic symmetry, we have $C_{11} = C_{22} =
C_{33}$, etc., which allows us to write $E_{\rm s}$ in a very compact
way.

Similarly, it is possible to show that the energy changes associated
to antiphase rotations are given by
\begin{equation}
\begin{split}
E_{r} = & \kappa_{r} r^{2} + \alpha_{r} r^{4}
+ \gamma_{r} (r_{x}^{2}r_{y}^{2} + r_{y}^{2}r_{z}^{2} +
r_{z}^{2}r_{x}^{2}) \; ,
\end{split}
\end{equation}
where $r = |{\bf r}|$ and ${\bf r} = (r_{x},r_{y},r_{z})$. Note that
the existence of antiphase rotational instabilities of the cubic
structure translates into a negative value of $\kappa_{r}$, which
requires us to consider fourth-order terms so that $E_{r}$ can be
bounded from below and the low-symmetry minima well defined. Note also
that the term associated to $\alpha_{r}$ only depends on the modulus
$r$, and is therefore isotropic; in contrast, $\gamma_{r}$ captures
the anisotropy energy, and its sign will determine the preferred
alignment of the antiphase rotation axis.

Interestingly, the expression for $E_{r}$ is formally identical to the
one corresponding to the energy change as a function of a three
dimensional polarization vector.\cite{kingsmith94} Further, it can be
shown that also $E_{m}$ has the same functional form; we have
\begin{equation}
\begin{split}
E_{m} = & \kappa_{m} m^{2} + \alpha_{m} m^{4}\\
& + \gamma_{m} (m_{x}^{2}m_{y}^{2} + m_{y}^{2}m_{z}^{2} +
m_{z}^{2}m_{x}^{2}) \; ,
\end{split}
\end{equation}
where $m = |{\bf m}|$ and ${\bf m} = (m_{x},m_{y},m_{z})$.

As regards the interactions between ${\bf r}$ and ${\bf m}$, we will
restrict ourselves to the lowest-order couplings, which are given by
\begin{equation}
E_{\rm int} = \alpha_{\rm int} r^{2}m^{2} + \beta_{\rm int}
(r_{x}^{2}m_{x}^{2} + r_{y}^{2}m_{y}^{2} + r_{z}^{2}m_{z}^{2}) \; .
\label{eq:int-energy}
\end{equation}
Note that this lowest-order interaction term is anharmonic; the
antiphase and in-phase rotations are decoupled at the harmonic level,
which is a direct consequence of their being associated to different
$q$-points.

Finally, we consider only the lowest-order terms of the interaction
between AFD patterns and strain, which are
\begin{equation}
\begin{split}
E_{sp} =
& B_{r1xx}(\eta_{1}r_{x}^{2}+\eta_{2}r_{y}^{2}+\eta_{3}r_{z}^{2})\\
&+B_{r1yy}[\eta_{1}(r_{y}^{2}+r_{z}^{2})+\eta_{2}(r_{z}^{2}+r_{x}^{2})
  +\eta_{3}(r_{x}^{2}+r_{y}^{2})] 
\\
&+B_{r4yz}(\eta_{4}r_{y}r_{z}+\eta_{5}r_{z}r_{x}+\eta_{6}r_{x}r_{y}) \\
& + B_{m1xx} (\eta_{1}m_{x}^{2}+\eta_{2}m_{y}^{2}+\eta_{3}m_{z}^{2})\\
& +
B_{m1yy}[\eta_{1}(m_{y}^{2}+m_{z}^{2})+\eta_{2}(m_{z}^{2}+m_{x}^{2})
  \\ & \;\;\;+\eta_{3}(m_{x}^{2}+m_{y}^{2})] \; .
\end{split}
\label{eq:energy-sp}
\end{equation}
Note that the form of the strain-phonon couplings is slightly different
for antiphase and in-phase tilts, as the former present a low-order
coupling with the shear strains while the latter do not. Indeed,
coupling terms of the type $\eta_{4}m_{y}m_{z}$ are not invariant under
the translations of the cubic lattice, which can be immediately seen
by noting that the $m_{y}$ and $m_{z}$ tilts are associated,
respectively to the $\pi/a (1,0,1)$ and $\pi/a (1,1,0)$ $q$-points,
while the shear strain is a zone-center distortion. (Some authors
include in the expression for the energy the coupling that we would
denote $B_{m4yz}$ in our notation;\cite{gu12} yet, such a coupling is
identically null by symmetry.)

Our expression for the PES of perovskite compounds with
O$_{6}$-rotational instabilities is thus complete. Note that, thanks
to the cubic symmetry of the reference structure, the list of
independent couplings controlling the behavior of these materials is
relatively short. We have three in $E_{\rm s}$, three in $E_{r}$,
three in $E_{m}$, two in $E_{\rm int}$, and five in $E_{\rm sp}$.

\subsection{Strain-renormalized energy function}
\label{sec:strain-free}

The cubic phase of simple {\sl AB}O$_{3}$ perovskites tends to be
stable against homogeneous strain deformations, so that $E_{\rm s}$ is
always positive. (More precisely, this implies that $C_{11}-C_{12}
>0$, $C_{11}+2C_{12} > 0$, and $C_{44}
>0$.\cite{born-book1954,karki97}) Hence, typically we can treat
strains as secondary distortions that simply follow the primary order
parameters ${\bf r}$ and ${\bf m}$. Mathematically, such equilibrium
strains $\{\eta^{\rm eq}_{a}\}$ can be obtained by imposing the
conditions
\begin{equation}
\left.\frac{\partial E}{\partial \eta_{a}}\right|_{\rm eq} = 0 \, ,
\end{equation}
for $a = 1, ..., 6$. These translate into the set of linear equations
\begin{equation}
\sum_{b} C_{ab} \eta^{\rm eq}_{b} = - \sum_{\alpha\beta}
B_{ra\alpha\beta} r_{\alpha}r_{\beta} - \sum_{\alpha\beta}
B_{ma\alpha\beta} m_{\alpha}m_{\beta} \; ,
\label{eq:system-of-eqs}
\end{equation}
which can be trivially resolved by inverting the $C_{ab}$ matrix:
\begin{equation}
  \eta^{\rm eq}_{a} = - \sum_{b} (C^{-1})_{ab} (B_{rb} + B_{mb})\; ,
\label{eq:equilibrium-strains}
\end{equation}
where
\begin{equation}
B_{rb} = \sum_{\alpha\beta} B_{rb\alpha\beta}
r_{\alpha}r_{\beta} 
\end{equation}
and
\begin{equation}
B_{mb} = \sum_{\alpha\beta} B_{mb\alpha\beta}
m_{\alpha}m_{\beta} \; .
\end{equation}
Without going into details, let us emphasize that the equilibrium
strains $\{\eta^{\rm eq}_{a}\}$ depend quadratically on the tilt
amplitudes. Hence, if substitute Eq.~(\ref{eq:equilibrium-strains}) in
Eq.~(\ref{eq:energy}), we obtain a simplified expression for a
strain-renormalized energy,
\begin{equation}
\begin{split}
  \bar{E}({\bf r},{\bf m}) = & E_{0} + \bar{E}_{r}({\bf r}) +
  \bar{E}_{m}({\bf m}) + \bar{E}_{\rm
    int}({\bf r},{\bf m}) \, ,
\end{split}
\label{eq:renorm-energy}
\end{equation}
where the barred energy terms are formally identical to the unbarred
ones described above, but contain modified anharmonic couplings. More
precisely, the strain terms in $E_{\rm s}$ will lead to renormalized
$\bar{\alpha}_{r}$ and $\bar{\gamma}_{r}$ interactions (coming from
the part of $\eta^{\rm eq}_{a}$ that is proportional to
$B_{ra\alpha\beta}$), renormalized $\bar{\alpha}_{m}$ and
$\bar{\gamma}_{m}$ couplings (coming from the part of $\eta^{\rm
  eq}_{a}$ that is proportional to $B_{ma\alpha\beta}$) and
renormalized $\bar{\alpha}_{\rm int}$ and $\bar{\beta}_{\rm int}$
couplings (coming from the crossed products between the ${\cal
  O}(r^{2})$ and ${\cal O}(m^{2})$ contributions to $\eta^{\rm
  eq}_{a}$). As for the $E_{\rm sp}$ term, it is linear in the strain
and quadratic in the rotation amplitudes; hence, by imposing $\eta_{a}
= \eta^{\rm eq}_{a}$, we again obtain additional contributions to the
fourth-order couplings in $E_{r}$, $E_{m}$, and $E_{\rm int}$. As a
result, the energies $\bar{E}_{r}$, $\bar{E}_{m}$, and $\bar{E}_{\rm
  int}$ in Eq.~(\ref{eq:renorm-energy}) have exactly the same
functional form as their respective counterparts in
Eq.~(\ref{eq:energy}), but with renormalized fourth-order couplings.

Note that it is possible to solve this problem analytically, as done
in the Appendix~A of Ref.~\onlinecite{kingsmith94} for an analogous
case. Let us also mention that the previous derivation is essentially
identical to the {\em stress-free} boundary conditions discussed in
the Appendix of Ref.~\onlinecite{gu12}, where explicit expressions for
the renormalized coefficients are given.

\subsection{Main singular points of the energy surface}
\label{sec:singular}

Let us now discuss the most important structures that may constitute
minima or saddle points of the renormalized energy in
Eq.~(\ref{eq:renorm-energy}). We leave strains out of the following
discussion for simplicity, noting that it is always possible to
obtain them from the rotation amplitudes by using
Eq.~(\ref{eq:equilibrium-strains}).

\subsubsection{Structures with antiphase rotations}

First, let us consider phases characterized by antiphase rotations
alone. As done in Ref.~\onlinecite{kingsmith94} for the formally
similar case of the electric polarization, let us distinguish three
different types of solutions.

{\sl $a^{-}b^{0}b^{0}$ structures}.-- We can have phases with ${\bf r}
= r(1,0,0)$, denoted $a^{-}b^{0}b^{0}$ in Glazer's notation. Note
that, equivalently, ${\bf r}$ could be (anti)parallel to the [010] or
[001] pseudo-cubic directions; hence, we have six symmetry-equivalent
states of this kind. Such structures have tetragonal symmetry with
space group $I4/mcm$, the low-temperature phase of SrTiO$_{3}$ being a
representative example. By substitution in
Eq.~(\ref{eq:renorm-energy}), we find that the energy of such a state
is given by
\begin{equation}
  \bar{E} = E_{0} + \kappa_{r} r^{2} + \bar{\alpha}_{r} r^{4}
  \; ,
\end{equation}
which can be minimized to render a singular point characterized by
\begin{equation}
  r^{\rm tet} = \left( - \frac{\kappa_{r}}{2\bar{\alpha}_{r}}
  \right)^{1/2}
\end{equation}
and
\begin{equation}
  E_{r}^{\rm tet} = E_{0} - \frac{\kappa_{r}^{2}}{4\bar{\alpha}_{r}}
  \; .
  \label{eq:Etet}
\end{equation}
Note that here we are assuming $\kappa_{r} < 0$, so that the antiphase
O$_{6}$ rotations constitute a structural instability of the cubic
perovskite phase. We also assume $\bar{\alpha}_{r}>0$, so that there
exists an optimum rotation amplitude that minimizes the
energy. Finally, note that we do not mark $E^{\rm tet}_{r}$ with a bar, as
this is the actual energy of the strain-relaxed $a^{-}b^{0}b^{0}$
state, i.e., it is exactly the same result we would obtain by working
with Eq.~(\ref{eq:energy}).

The stability of this solution can be readily analyzed by computing
the Hessian matrix for $\bar{E}$ at ${\bf r} = {\bf r}^{\rm tet}$ and
${\bf m} = {\bf 0}$. Let us consider states given by ${\bf r} = r^{\rm
  tet}(1,0,0) + \delta {\bf r}$ and ${\bf m} = \delta {\bf
  m}$. The 6-dimensional Hessian associated to the $\delta{\bf r}$ and
$\delta{\bf m}$ perturbations has the diagonal form
\begin{equation}
H^{\rm tet} =
\begin{bmatrix}
\kappa_{r\parallel}^{\rm tet} & 0 & 0 & 0 & 0 & 0\\
0 & \kappa_{r\perp}^{\rm tet} & 0 & 0 & 0 & 0 \\
0 & 0 & \kappa_{r\perp}^{\rm tet} & 0 & 0 & 0  \\
0 & 0 & 0 & \kappa_{m\parallel}^{\rm tet} & 0 & 0 \\
0 & 0 & 0 & 0 & \kappa_{m\perp}^{\rm tet} & 0 \\
0 & 0 & 0 & 0 & 0 & \kappa_{m\perp}^{\rm tet}
\end{bmatrix} \; ,
\end{equation}
where
\begin{equation}
\begin{split}
  \kappa_{r\parallel}^{\rm tet} = & -4\kappa_{r} \\
  \kappa_{r\perp}^{\rm tet} =    & -\kappa_{r}\frac{\bar{\gamma}_{r}}{\bar{\alpha}_{r}}
  \\
  \kappa_{m\parallel}^{\rm tet} = &
  2\left(\kappa_{m}-\kappa_{r}\frac{\bar{\alpha}_{\rm
        int}+\bar{\beta}_{\rm int}}{2\bar{\alpha}_{r}}\right) \\
  \kappa_{m\perp}^{\rm tet} = &
  2\left(\kappa_{m}-\kappa_{r}\frac{\bar{\alpha}_{\rm int}}{2\bar{\alpha}_{r}}\right)
  \; .
\end{split}
\end{equation}
From these results, a few interesting conclusions immediately
follow. As regards the antiphase rotations, we naturally have that the
${\bf r} = {\bf r}^{\rm tet}$ state is stable against parallel
perturbations, since $\kappa_{r\parallel}^{\rm tet} > 0$. In contrast,
the stability with respect to transversal perturbations depends on the
sign of $\bar{\gamma}_{r}$: a positive value indicates that the
tetragonal phase is stable against such distortions
($\kappa_{r\perp}^{\rm tet} > 0$), but a negative $\bar{\gamma}_{r}$
implies we have a saddle point ($\kappa_{r\perp}^{\rm tet} < 0$).

As regards the in-phase rotations, the ${\bf r} = {\bf r}^{\rm tet}$
solution will be stable against them whenever we have a large enough
$\kappa_{m} > 0$. A more interesting (and more typical) situation
occurs if $\kappa_{m}<0$, i.e., whenever the in-phase tilts are
instabilities of the cubic phase. In such a case, the sign of
$\kappa_{m\perp}^{\rm tet}$ will be positive provided that
\begin{equation}
|\kappa_{m}| < |\kappa_{r}|\frac{\bar{\alpha}_{\rm
    int}}{2\bar{\alpha}_{r}} \; .
\end{equation}
This would be a situation in which the competition between antiphase
and in-phase rotations, as quantified by the coupling
$\bar{\alpha}_{\rm int} > 0$, is large enough so that the presence of
the former prevents the occurrence of the latter. Note that
$\bar{\alpha}_{\rm int}$ only accounts for an isotropic competition
between different tilt types, while $\bar{\beta}_{\rm int}$ -- which
appears in $\kappa_{m\parallel}^{\rm tet}$ but not in
$\kappa_{m\perp}^{\rm tet}$ -- also includes a directional
contribution.

Finally, note that for $\bar{\alpha}_{\rm int} < 0$ or
$\bar{\alpha}_{\rm int}+\bar{\beta}_{\rm int} < 0$ we would have a
cooperation between different rotational patterns. In such a case the
$a^{-}b^{0}b^{0}$ state would not be an energy minimum, unless the
in-phase tilt is robustly stable (i.e., $\kappa_{m} > 0$ and large
enough).

{\sl $a^{-}a^{-}a^{-}$ structures}.-- Another important case
corresponds to rhombohedral structures like those of LaAlO$_{3}$ and
LaNiO$_{3}$, with space group $R\bar{3}c$, displaying antiphase
rotations of equal amplitude about the three pseudo-cubic
axes. Equivalently, we can think of a single rotation about
[111]. (Note that there are eight symmetry-equivalent states of this
type.) The corresponding singular point is given by ${\bf r} = {\bf
  r}^{\rm rho} = r^{\rm rho}(1,1,1)$ and ${\bf m} = {\bf 0}$ with
\begin{equation}
  r^{\rm rho} = \left( -
  \frac{\kappa_{r}}{6\bar{\alpha}_{r}+2\bar{\gamma}_{r}}
  \right)^{1/2}
\end{equation}
and
\begin{equation}
  E_{r}^{\rm rho} = E_{0} -
  \frac{\kappa_{r}^{2}}{4\left(\bar{\alpha}_{r}
    +\frac{1}{3}\bar{\gamma}_{r}\right)}
  \; .
  \label{eq:Erho}
\end{equation}
Here we assume that $\kappa_{r} < 0$ and $\bar{\alpha}_{r} +
2\bar{\gamma}_{r} > 0$ (with $\bar{\alpha}_{r} > 0$, as mentioned
above), so that $r^{\rm rho}$ is well defined. As above, we can study
the stability of this solution by computing the corresponding Hessian
matrix. We thus consider states given by ${\bf r} = r^{\rm
  rho}(1,1,1) + \delta {\bf r}$ and ${\bf m} = \delta {\bf m}$,
and work with symmetry-adapted distortions so that
\begin{equation}
\begin{split}
\delta {\bf r} = & \frac{\delta r_{\parallel}}{\sqrt{3}}  (1,1,1) +
\frac{\delta r_{\perp1}}{\sqrt{2}}  (1,\bar{1},0) +
\frac{\delta r_{\perp2}}{\sqrt{6}}  (1,1,\bar{2}) \\
\delta {\bf m} = & \frac{\delta m_{\parallel}}{\sqrt{3}}  (1,1,1) +
\frac{\delta m_{\perp1}}{\sqrt{2}}  (1,\bar{1},0) +
\frac{\delta m_{\perp2}}{\sqrt{6}}  (1,1,\bar{2}) \; .
\end{split}
\end{equation}
It can be proved that the Hessian matrix is diagonal in this basis; we
have 
\begin{equation}
H^{\rm rho} =
\begin{bmatrix}
\kappa_{r\parallel}^{\rm rho} & 0 & 0 & 0 & 0 & 0\\
0 & \kappa_{r\perp}^{\rm rho} & 0 & 0 & 0 & 0 \\
0 & 0 & \kappa_{r\perp}^{\rm rho} & 0 & 0 & 0  \\
0 & 0 & 0 & \kappa_{m}^{\rm rho} & 0 & 0 \\
0 & 0 & 0 & 0 & \kappa_{m}^{\rm rho} & 0 \\
0 & 0 & 0 & 0 & 0 & \kappa_{m}^{\rm rho}
\end{bmatrix} \; ,
\end{equation}
where
\begin{equation}
\begin{split}
  \kappa_{r\parallel}^{\rm rho} = & -4\kappa_{r} \; , \\
  \kappa_{r\perp}^{\rm rho} = & \;\;
  2\kappa_{r} \frac{\bar{\gamma_r}}{3\bar{\alpha}_{r}+\bar{\gamma}_{r}}
  \; , \\
  \kappa_{m}^{\rm rho} = & \;\;
  2\left(\kappa_{m}-\kappa_{r}\frac{3\bar{\alpha}_{\rm
        int}+\bar{\beta}_{\rm
        int}}{6\bar{\alpha}_{r}+2\bar{\gamma}_{\rm int}}\right)
  \; .
\end{split}
\end{equation}
This result bears obvious similarities with what we obtained above for
the $a^{-}b^{0}b^{0}$ state. One interesting observation is that
$\kappa_{r\perp}^{\rm rho}$ and $\kappa_{r\perp}^{\rm tet}$ must
necessarily have opposite signs. This implies that, in our
fourth-order PES, the $a^{-}b^{0}b^{0}$ and $a^{-}a^{-}a^{-}$ states
cannot be energy minima simultaneously, and their relative stability
is controlled by the sign of the $\bar{\gamma}_{r}$ parameter. Note
that this observation is consistent with the discussion in
Ref.~\onlinecite{vanderbilt01} on the conditions for having multiple
stable states in potential energy surfaces of the same type as
$\bar{E}_{r}$.

{\sl $a^{-}a^{-}c^{0}$ structures}.-- Finally, the last structure of
this series is the orthorhombic phase with space group $Imcm$ and
antiphase rotations of equal amplitude about two pseudo-cubic
axes. (This amounts to a rotation about a $\langle 110 \rangle$
direction. Note that there are twelve symmetry-equivalent structures
of this type.) The corresponding stationary point is defined by ${\bf
  r} = {\bf r}^{\rm ort} = r^{\rm ort}(1,1,0)$ and ${\bf m} = {\bf 0}$
with
\begin{equation}
  r^{\rm ort} = \left( -
  \frac{\kappa_{r}}{4\bar{\alpha}_{r}+\bar{\gamma}_{r}} \right)^{1/2}
\end{equation}
and
\begin{equation}
  E_{r}^{\rm ort} = E_{0} -
  \frac{\kappa_{r}^{2}}{4{\alpha}_{r}
    +\bar{\gamma}_{r}} \; ,
  \label{eq:Eort}
\end{equation}
where, as above, we assume that $\kappa_{r} < 0$ and
$4\bar{\alpha}_{r} + \bar{\gamma}_{r} > 0$ with $\bar{\alpha}_{r} >
0$, so that $r^{\rm ort}$ is a real number. To study the stability of
the solution, we consider states given by ${\bf r} = r^{\rm
  ort}(1,1,0) + \delta {\bf r}$ and ${\bf m} = \delta {\bf m}$,
and work with symmetry-adapted distortions defined as
\begin{equation}
\begin{split}
\delta {\bf r} = & \frac{\delta r_{\parallel}}{\sqrt{2}}  (1,1,0) +
\frac{\delta r_{\perp1}}{\sqrt{2}}  (1,\bar{1},0) +
\delta r_{\perp2} (0,0,1) \\
\delta {\bf m} = & \frac{\delta m_{\parallel}}{\sqrt{2}}  (1,1,0) +
\frac{\delta m_{\perp1}}{\sqrt{2}}  (1,\bar{1},0) +
\delta m_{\perp2} (0,0,1) \\
\end{split}
\end{equation}
In this basis, the Hessian is diagonal and has the form
\begin{equation}
H^{\rm ort} =
\begin{bmatrix}
\kappa_{r\parallel}^{\rm ort} & 0 & 0 & 0 & 0 & 0\\
0 & \kappa_{r\perp1}^{\rm ort} & 0 & 0 & 0 & 0 \\
0 & 0 & \kappa_{r\perp2}^{\rm ort} & 0 & 0 & 0  \\
0 & 0 & 0 & \kappa_{m\parallel}^{\rm ort} & 0 & 0 \\
0 & 0 & 0 & 0 & \kappa_{m\perp1}^{\rm ort} & 0 \\
0 & 0 & 0 & 0 & 0 & \kappa_{m\perp2}^{\rm ort}
\end{bmatrix} \; ,
\end{equation}
where
\begin{equation}
\begin{split}
  \kappa_{r\parallel}^{\rm ort} = & -4\kappa_{r} \; , \\
  \kappa_{r\perp1}^{\rm ort} = & \;\;
  6\kappa_{r} \frac{\bar{\gamma_r}}{4\bar{\alpha}_{r}+\bar{\gamma}_{r}}
  \; , \\
  \kappa_{r\perp2}^{\rm ort} = & \;\;
  -2\kappa_{r} \frac{\bar{\gamma_r}}{4\bar{\alpha}_{r}+\bar{\gamma}_{r}}
  \; , \\
  \kappa_{m\parallel}^{\rm ort} = &
  2\left(\kappa_{m}-\kappa_{r}\frac{2\bar{\alpha}_{\rm
      int}+\bar{\beta}_{\rm
      int}}{4\bar{\alpha}_{r}+\bar{\gamma}_{r}}\right) \; , \\
  \kappa_{m\perp1}^{\rm ort} = &
  2\left(\kappa_{m}-\kappa_{r}\frac{2\bar{\alpha}_{\rm
      int}+\bar{\beta}_{\rm
      int}}{4\bar{\alpha}_{r}+\bar{\gamma}_{r}}\right)
  \; , \\
  \kappa_{m\perp2}^{\rm ort} = &
  2\left(\kappa_{m}-\kappa_{r}\frac{2\bar{\alpha}_{\rm
      int}}{4\bar{\alpha}_{r}+\bar{\gamma}_{r}}\right)
  \; .
\end{split}
\end{equation}
As compared to the two cases above, the main peculiarity of this
result lies on the fact that $\kappa^{\rm ort}_{r\perp1}$ and
$\kappa^{\rm ort}_{r\perp2}$ must necessarily have opposite signs,
which implies that the orthorhombic $a^{-}a^{-}c^{0}$ solution cannot
be a minimum of the energy in our fourth-order PES.

\subsubsection{Structures with in-phase rotations}

As regards the states in which only in-phase rotations are condensed
-- denoted by $a^{+}b^{0}b^{0}$, $a^{+}a^{+}b^{0}$, and
$a^{+}a^{+}a^{+}$, respectively --, the situation is exactly analogous
to that of the purely antiphase structures. Indeed, since
$\bar{E}({\bf r})$ and $\bar{E}({\bf m})$ have the same functional
form, our previous discussion can be directly applied to the phases
with pure in-phase tilts by simply making the substitution $r
\rightarrow m$.

\subsubsection{The $a^{-}a^{-}c^{+}$ structure}

Finally, let us discuss the case of the orthorhombic $Pbnm$ phase that
combines antiphase tilts about [110] with in-phase tilts about
[001]. (There are twenty-four symmetry-equivalent structures of this
type.) In this case the distortion has the form ${\bf r} = r (1,1,0)$
and ${\bf m} = m (0,0,1)$, and the energy is
\begin{equation}
\begin{split}
  \bar{E} = & \; E_{0} + 2 \kappa_{r} r^{2} +
  (4\bar{\alpha}_{r}+\bar{\gamma}_{r}) r^{4} \\
    & \; + \kappa_{m} m^{2} + \bar{\alpha}_{m} m^{4} +
  2\bar{\alpha}_{\rm int}r^{2}m^{2} \; .
\end{split}
\label{eq:energy-ophase-raw}
\end{equation}
For simplicity, in the following we use $\bar{\alpha}^{*} =
4\bar{\alpha}+\bar{\gamma}$. Minimizing this energy renders a
structure given by
\begin{equation}
(r^{\cal O})^{2} =
  \frac{-\kappa_{r}}{\bar{\alpha}^{*}_{r}-\bar{\alpha}_{\rm
      int}^{2}/\bar{\alpha}_{m}}
+ \frac{\bar{\alpha}_{\rm
    int}\kappa_{m}}{2(\bar{\alpha}^{*}_{r}\bar{\alpha}_{m} -
  \bar{\alpha}_{\rm int}^{2})}
\end{equation}
and
\begin{equation}
(m^{\cal O})^{2} =
  \frac{-\kappa_{m}}{2(\bar{\alpha}_{m}-\bar{\alpha}_{\rm
      int}^{2}/\bar{\alpha}^{*}_{r})}
+ \frac{\bar{\alpha}_{\rm
    int}\kappa_{r}}{\bar{\alpha}^{*}_{r}\bar{\alpha}_{m} -
  \bar{\alpha}_{\rm int}^{2}} \; ,
\label{eq:ophase-m}
\end{equation}
where we use the notation ${\cal O}$ to distinguish this orthorhombic
($a^{-}a^{-}c^{+}$) phase from the simpler ``ort'' tilt patterns
($a^{-}a^{-}c^{0}$ and $a^{+}a^{+}c^{0}$) discussed above. The energy
for this state is
\begin{equation}
\begin{split}
  E^{\cal O} = & \; E_{0} - \frac{\kappa_{r}^{2}}{\bar{\alpha}^{*}_{r}
    - \bar{\alpha}_{\rm int}^{2}/\bar{\alpha}_{m}} -
  \frac{\kappa_{m}^{2}}{4(\bar{\alpha}_{m} - \bar{\alpha}_{\rm
      int}^{2}/\bar{\alpha}^{*}_{r})} \\
& \; + \frac{\kappa_{r}\kappa_{m}\bar{\alpha}_{\rm
      int}}{\bar{\alpha}^{*}_{r}\bar{\alpha_{m}}-\bar{\alpha}_{\rm
      int}^{2}} \; .
\end{split}
\label{eq:Eo}
\end{equation}
From the previous expressions, it is obvious that in absence of
interaction between antiphase and in-phase rotations -- i.e., for
$\bar{\alpha}_{\rm int} = 0$ -- the ${\cal O}$ phase reduces to a
trivial combination of the orthorhombic ${\bf r} = {\bf r}^{\rm ort}$
and tetragonal ${\bf m} = {\bf m}^{\rm tet}$ states described
above. It is also possible to prove that, for the ${\cal O}$ solution
to exist, at least one of the rotational modes must be an instability
of the cubic phase, i.e., either $\kappa_{r}$ or $\kappa_{m}$, or both
of them, must be negative. Indeed, if we have $\kappa_{r}, \kappa_{m}
> 0$ -- and given that $\bar{\alpha}^{*}_{r} , \bar{\alpha}_{m} >0$ as
required for the energy to be bounded from below --, there is no
choice of $\bar{\alpha}_{\rm int}$ that can yield a well-defined
${\cal O}$ state.

To study the stability of this solution, we consider the structures
given by ${\bf r} = {\bf r}^{\cal O} + \delta{\bf r}$ and ${\bf m} =
{\bf m}^{\cal O} + \delta{\bf m}$, and work with symmetry-adapted
distortions defined by
\begin{equation}
\begin{split}
\delta {\bf r} = & \frac{\delta r_{\parallel1}}{\sqrt{2}}  (1,1,0) +
\delta r_{\parallel2} (0,0,1) +
\frac{\delta r_{\perp}}{\sqrt{2}}  (1,\bar{1},0) \\
\delta {\bf m} = & \frac{\delta m_{\parallel1}}{\sqrt{2}}  (1,1,0) +
\delta m_{\parallel2} (0,0,1) +
\frac{\delta m_{\perp}}{\sqrt{2}}  (1,\bar{1},0) \; .
\end{split}
\end{equation}
In this basis, the Hessian has the form
\begin{equation}
H^{\cal O} =
\begin{bmatrix}
\kappa_{r\parallel1}^{\cal O} & 0 & 0 & 0 & \kappa_{rm\parallel}^{\cal
O} & 0\\
0 & \kappa_{r\parallel2}^{\cal O} & 0 & 0 & 0 & 0 \\
0 & 0 & \kappa_{r\perp}^{\cal O} & 0 & 0 & 0  \\
0 & 0 & 0 & \kappa_{m\parallel1}^{\cal O} & 0 & 0 \\
\kappa_{rm\parallel}^{\cal O} & 0 & 0 & 0 & \kappa_{m\parallel2}^{\cal
  O} & 0 \\
0 & 0 & 0 & 0 & 0 & \kappa_{m\perp}^{\cal O}
\end{bmatrix} \; ,
\end{equation}
where
\begin{equation}
\begin{split}
  \kappa_{r\parallel1}^{\cal O} = \; & 2\kappa_{r} + 6 (r^{\cal O})^2
  \bar{\alpha}^{*}_{r} + 2 (m^{\cal O})^2 \bar{\alpha}_{\rm int}
  \; , \\
  \kappa_{r\parallel2}^{\cal O} = \; & 2\kappa_{r} + 4 (r^{\cal O})^2
  (2\bar{\alpha}_{r}+\bar{\gamma}_{r}) + 2 (m^{\cal O})^2 \bar{\alpha}_{\rm int}
  \; , \\
  \kappa_{r\perp}^{\cal O} = \; & 2\kappa_{r} + 4 (r^{\cal O})^2
  (2\bar{\alpha}_{r}-\bar{\gamma}_{r}) \\
  & \;\;\;\;\;\;\; + 2 (m^{\cal O})^2
  (\bar{\alpha}_{\rm int}+\bar{\beta}_{\rm int})
  \; , \\
  \kappa_{m\parallel1}^{\cal O} = \; & 2\kappa_{m} + 2 (m^{\cal O})^2
  (2\bar{\alpha}_{m}+\bar{\gamma}_{m}) \\
  & \;\;\;\;\;\;\; + 2 (r^{\cal O})^2
  (2\bar{\alpha}_{\rm int}+\bar{\beta}_{\rm int})
  \; , \\
  \kappa_{m\parallel2}^{\cal O} = \;  & 2\kappa_{m} + 12 (m^{\cal O})^2
  \bar{\alpha}_{m} + 4 (r^{\cal O})^2 \bar{\alpha}_{\rm int}
  \; , \\
  \kappa_{m\perp}^{\cal O} = \;  & 2\kappa_{m} + 2 (m^{\cal O})^2
  (2\bar{\alpha}_{m}+\bar{\gamma}_{m}) \\
  & \;\;\;\;\;\;\; + 2 (r^{\cal O})^2
  (2\bar{\alpha}_{\rm int}+\bar{\beta}_{\rm int})
  \; , \\
  \kappa_{rm\parallel}^{\cal O} = \;  & 4\sqrt{2} r^{\cal O} m^{\cal O}
  \bar{\alpha}_{\rm int} \; .
\end{split}
\label{eq:hessian-elements-ophase}
\end{equation}
Note that, at variance with the Hessian matrices introduced above,
this one is not diagonal. Indeed, because both $\delta r_{\parallel
  1}$ and $\delta m_{\parallel 2}$ correspond to fully symmetric
distortions in the ${\cal O}$ phase, there is a non-zero off-diagonal
coupling between them. Naturally, by construction, the ${\cal O}$
phase is stable against such distortions, and this part of the Hessian
is positive definite. The stability of the ${\cal O}$ phase thus
relies on the parameters $\kappa_{r\parallel 2}^{\cal O}$,
$\kappa_{m\parallel 1}^{\cal O}$, $\kappa_{r\perp}^{\cal O}$, and
$\kappa_{m\perp}^{\cal O}$, which should all be positive. It is clear
that this will depend on the relative values of the $\bar{\alpha}$ and
$\bar{\gamma}$ anharmonic couplings affecting individual tilt patterns
(e.g., a positive $\bar{\gamma}_{m}$ favoring the $a^{+}b^{0}b^{0}$
configuration over $a^{+}a^{+}a^{+}$ will obviously be helpful), and
on the strength of the competing/cooperative interactions between
different tilt types (thus, e.g., a positive $\bar{\beta}_{\rm int}$
will be generally beneficial for the stability of the ${\cal O}$
phase). It is worth noting that the existence of the $a^{-}a^{-}c^{+}$
pattern as the ground state also requires that its energy $E^{\cal O}$
be lower than that of competing polymorphs ($E_{r}^{\rm rho}$, etc.),
which imposes additional conditions on the coupling parameters.

We will not analyze here all the possibilities and parameter
combinations that may result in the stabilization of the $Pbnm$
phase. Instead, below we will focus on discussing the parameter values
that are typical of actual materials displaying the $a^{-}a^{-}c^{+}$
ground state. As we will see, all the investigated $Pbnm$ compounds
present a rather similar behavior, and a very clear physical picture
emerges.

\section{Computational approach}
\label{sec:methods}

We use first-principles simulation methods to investigate thirty-five
representative perovskite oxides with low-energy structures
characterized by O$_{6}$ rotations. The chosen compounds tend to have
small tolerance factors ranging between $t = 0.81$ (ZnSnO$_{3}$) and
$t = 1.00$ (BaZrO$_{3}$). In some cases, we consider various members
of significant materials families -- as e.g. for the ${\sl
  A}$FeO$_{3}$ orthoferrites --, so that trends as a function of $t$
can be more clearly identified. Note that all the considered compounds
are simple {\sl AB}O$_{3}$ perovskites with uniquely defined {\sl A}
and {\sl B} cations. Nevertheless, since the structural
  properties of perovskite solid solutions of the form ({\sl A},{\sl
    A}')({\sl B},{\sl B}')O$_{3}$ tend to depend smoothly on
  composition (Vegard's law),\cite{denton91,bellaiche00b,iniguez03}
we believe that our conclusions should be applicable to such more
complex compounds, at least as regards trends dominated by cation size
or steric effects.

Note that some of the considered compounds -- especially small-$t$
ones -- may display (anti)polar instabilities of their cubic phase, in
addition to the AFD soft modes. In such cases, a complete PES model
should include, on top of the description of the tilting modes, an
explicit theory of the most important polar order parameters, which
would complicate the treatment considerably and remains for future
work. Here, all such degrees of freedom are treated implicitly, and
they are assumed to follow the primary AFD order parameters in what
concerns the discussion of the tilted structures. We should note that,
in a few cases, the actual ground state of such materials may be FE,
or may combine FE and AFD distortions. However, for the purpose of the
present discussion, we will only consider structures in which the AFD
modes are the primary order. For example, ZnSnO$_{3}$ has the $R3c$
ground state structure that is typical of
LiNbO$_{3}$;\cite{inaguma08,benedek13,gu17} further, ZnTiO$_{3}$ and
ZnGeO$_{3}$ have an ilmenite-type ground state.\cite{inaguma14,ross10}
The present discussion does not consider such structures and, thus, is
not intended to be a complete investigation of these compounds. Yet,
we include them among our studied materials, as they provide us with
valuable information on the behavior for very small tolerance factors.

To obtain information about the PES, we run symmetry-constrained
structural relaxations corresponding to the following tilt systems:
$a^{-}a^{-}c^{+}$, $a^{-}a^{-}a^{-}$, $a^{-}b^{0}b^{0}$,
$a^{+}a^{+}a^{+}$, and $a^{+}b^{0}b^{0}$. We also optimize the cubic
structure to obtain the reference energy $E_{0}$, and calculate the
elastic constants $C_{ab}$ from the response of this phase to small
strains. Further, we run structural relaxations under several
constraints -- e.g., by imposing the cell optimized for the cubic
structure (i.e., $\eta_{a} = 0$ $\forall a$), by disallowing the
off-centering displacements of the {\sl A} cations -- to further test
the behavior of the investigated materials. Such especial situations
are described in detail below.

We fit the {\em bare} coupling parameters
[Eqs.~(\ref{eq:energy})--(\ref{eq:energy-sp})] by imposing that our
models reproduce the AFD amplitudes, strains, and energies obtained
for the relaxed structures. More precisely, the $\kappa_{r}$,
$\alpha_{r}$, and $\gamma_{r}$ parameters are obtained so as to
reproduce exactly $E_{r}^{\rm rho}$ and $E_{r}^{\rm tet}$, as well as
the zero-derivative condition at the relaxed $a^{-}b^{0}b^{0}$
state. Similarly, $\kappa_{m}$, $\alpha_{m}$, and $\gamma_{m}$ are
fitted to reproduce $E_{m}^{\rm rho}$, $E_{m}^{\rm tet}$, and the
zero-derivative condition at the $a^{+}b^{0}b^{0}$ phase. The
$B_{1rxx}$, $B_{1ryy}$, and $B_{4ryz}$ couplings are obtained by
fitting the the $\eta_{1}$-derivatives of the energy evaluated at the
$a^{-}a^{-}a^{-}$ and $a^{-}b^{0}b^{0}$ phases, as well as the
$\eta_{4}$ derivative of the energy for the $a^{-}a^{-}a^{-}$
structure. Similarly, $B_{1mxx}$ and $B_{1myy}$ are obtained from the
$\eta_{1}$-derivatives of the energy evaluated at the
$a^{+}a^{+}a^{+}$ and $a^{+}b^{0}b^{0}$ structures. Finally, we fit
$\alpha_{\rm int}$ so that we reproduce the energy and zero-derivative
conditions of the ${\cal O}$ phase as well as possible. Additionally,
we consider a $a^{-}a^{-}a^{-}$ structure which we distort by hand,
imposing a small in-phase rotation about the [100] pseudo-cubic axis,
as needed to compute the coupling $\beta_{\rm int}$.

As for the strain-renormalized parameters, we follow essentially the
same procedure as above, demanding that the energy given by
Eq.~(\ref{eq:renorm-energy}) reproduces all the features of the
relaxed stationary structures, except the strains.

We find that the assumed fourth-order polynomial energy is sufficient
to obtain a satisfactory description of the key polymorphs mentioned
above for all the materials considered. Most importantly, the
interaction parameters $\alpha_{\rm int}$ and $\bar{\alpha}_{\rm int}$
are sufficient to capture the key ${\bf r}$-${\bf m}$ coupling, and
our models yield $E^{\cal O}-E_{0}$ values that deviate from the
first-principles result by about 2~\%, typically. Then, as we will see
in Section~\ref{sec:remarks}, obtaining a quantitatively (very)
accurate description of additional polymorphs (e.g.,
$a^{+}b^{-}a^{+}$) may require consideration of higher-order
interaction terms; however, this detail is not relevant for our
present discussion.

For the first-principles calculations, we use density functional
theory\cite{kohn65,hohenberg64} (DFT) within the generalized gradient
approximation adapted for solids (the so-called
``PBEsol''),\cite{perdew08} as implemented in the simulation package
{\sc VASP}.\cite{kresse96,kresse99} In the case of the considered
ferrites, we use a Hubbard-$U$ correction of the energy functional,
for a better description of iron's 3$d$ electrons,\cite{dudarev98}
choosing $U_{\rm eff} = 3.8$~eV which is known to work well for these
compounds;\cite{zhao16,kornev07,dieguez11} we also assume the iron
spins are in an anti-ferromagnetic arrangement, with antiparallel
first-nearest neighbors, mimicking their well-known ground-state
magnetic structure.\cite{white69} For SrRuO$_{3}$ and LaNiO$_{3}$, we
do not use any Hubbard-$U$ correction, and consider a trivial
ferromagnetic spin arrangement as starting point of our simulations;
for SrRuO$_{3}$ this yields the magnetic solution that has been
obtained in previous DFT investigations of this compound, and
basically coincides with the experimental state;\cite{miao14} for
LaNiO$_{3}$ our simulations yield a non-magnetic configuration, thus
reproducing previous calculations and agreeing well with the
experimental result.\cite{gibert12,weber16} Nevertheless,
  one should keep in mind that the adecuacy of a simple DFT treatment
  is questionable for such challenging compunds and, hence, our
  quantitative results for SrRuO$_{3}$ and LaNiO$_{3}$ should be
  regarded with some caution. The interaction between core and
valence electrons is treated using the projector augmented wave (PAW)
method,\cite{blochl94} solving explicitly for the following electrons:
O's 2$s$ and 2$p$; Na's 2$s$, 2$p$, and 3$s$; Al's 3$s$ and 3$p$; Ca's
3$s$, 3$p$, and 4$s$; Ti's 3$p$, 4$s$, and 3$d$; Cr's 3$p$, 4$s$, and
3$d$; Fe's 3$p$, 4$s$, and 3$d$; Ni's 3$p$, 4$s$, and 3$d$; Zn's 4$s$
and 3$d$; Ga's 4$s$, 3$d$, and 4$p$; Ge's 4$s$, 3$d$, and 4$p$; Sr's
4$s$, 4$p$, and 5$s$; Y's 4$s$, 4$p$, 5$s$, and 4$d$; Zr's 4$s$, 4$p$,
5$s$, and 4$d$; Ru's 4$s$, 4$p$, 5$s$, and 4$d$; Sn's 5$s$ and 5$p$;
Ba's 5$s$, 5$p$, and 6$s$; La's 5$s$, 5$p$, 6$s$, and 5$d$; Pr's 5$s$,
6$s$, 5$p$, and 5$d$; Nd's 5$s$, 6$s$, 5$p$, and 5$d$; Sm's 5$s$,
6$s$, 5$p$, and 5$d$; Gd's 6$s$, 5$p$, and 5$d$; Dy's 6$s$, 5$p$, and
5$d$; Yb's 6$s$, 5$p$, and 5$d$; Hf's 5$p$, 6$s$, and 5$d$; Ta's 6$s$
and 5$d$. Electronic wave functions are described in a plane wave
basis cut off at 500~eV. All the investigated structures are treated
using the same 40-atom {\em Glazer cell}, which can be viewed as a
$2\times2\times2$ multiple of the elemental 5-atom perovskite unit
and is compatible with all the AFD patterns of interest
here. Brillouin zone integrals corresponding to this cell are computed
using a $\Gamma$-centered $3\times3\times3$ grid of $k$-points. (Note
that except for SrRuO$_{3}$ and LaNiO$_{3}$ -- for which a grid of
$9\times9\times9$ $k$-points is used -- all the considered materials
are insulators.)  Structural relaxations are stopped when residual
forces and stresses are below 0.01~eV/\AA\ and 0.2~GPa,
respectively. We checked that these calculation conditions are
well-converged and sufficient for our current purposes.

 Our results are in reasonable agreement with previous
  first-principles calculations in the literature. Representative of
  this are the elastic constants, for which there is plenty of
  published data for some compounds. For example, for CaTiO$_{3}$ we
  obtain $C_{11} = 373$~GPa, $C_{12} = 103$~GPa, and $C_{44} = 99$~GPa
  from our PBEsol calculations (see
  Table~\ref{tab:bare-parameters}). In contrast, a work \cite{long13}
  based on a different generalized-gradient approximation
  \cite{perdew96} reports values of 331~GPa, 96~GPa, and 95~GPa,
  respectively; while the authors of Ref.~\onlinecite{gu12} obtained
  403~GPa, 107~GPa, and 100~GPa, respectively, when using a
  local-density approximation\cite{kohn65} to DFT. Hence, our
  numerical results fall within the accuracy that can be expected from
  first-principles calculations that, besides other technical details,
  depend significantly on the choice of density functional.

Finally, let us mention some important details for the calculation of
structural parameters and coupling constants. As mentioned above,
${\bf r}$ and ${\bf m}$ are the amplitudes of the antiphase and
in-phase AFD order parameters, respectively (see
Fig.~\ref{fig:sketch-dists}). Then, let $\{r_{l\kappa\alpha}\}$ be the
atomic positions corresponding to an arbitrary configuration of our
periodically-repeated Glazer cell; here, $l$ labels the individual
5-atom cells inside our 40-atom supercell, $\{R_{l\beta}\}$ being the
corresponding lattice vectors; $\kappa$ labels the atoms inside a
5-atom cell, whose positions in the cubic reference structure are
given by $\tau_{\kappa\beta}$; $\alpha$ and $\beta$ label the
Cartesian axes, which coincide with the pseudo-cubic directions of the
perovskite structure. Then, such a configuration can be expressed as
\begin{equation}
r_{l\kappa\alpha} = \sum_{\beta} (\delta_{\alpha\beta} +
\eta_{\alpha\beta})(R_{l\beta} + \tau_{\kappa\beta}) +
u_{l\kappa\alpha} \; ,
\label{eq:distortions}
\end{equation}
where we have written the strains $\eta_{\alpha\beta}$ in their full
tensor form, avoiding the compact Voigt notation. More importantly,
Eq.~(\ref{eq:distortions}) introduces the quantities
$\{u_{l\kappa\alpha}\}$, i.e., the atomic distortions with respect to
the strained reference structure. From these distortions, we obtain
the amplitudes ${\bf r}$ and ${\bf m}$ by projecting onto six
symmetry-adapted modes associated to each of the three antiphase and
three in-phase octahedral rotations. We use modes that are normalized
to unity when we sum over atoms in the 40-atom cell. The resulting
amplitudes ${\bf r}$ and ${\bf m}$ thus have units of length (we use
\AA\ throughout). Hence, the harmonic constants $\kappa_{r}$ and
$\kappa_{m}$ in our energy function are given in
eV/\AA$^{2}$, the 4th-order couplings ($\alpha_{r}$,
  $\beta_{r}$, $\alpha_{\rm int}$, etc.) are in eV/\AA$^{4}$, and the
  6th-order correction $\bar{\gamma}_{\rm int}$ in eV/\AA$^{6}$. As
usual, the strains are adimensional, so that the elastic
  constants are given in eV and the strain-phonon couplings
  ($B_{1rxx}$, etc.) in eV/\AA$^{2}$. Finally, note that
all the parameters are normalized so that the functions
$E({\bf r},{\bf m},\{\eta_{a}\})$ and $\bar{E}({\bf r},{\bf m})$ give
energy per 40-atom cell.

\begin{figure*}
  \includegraphics[width=1.00\linewidth]{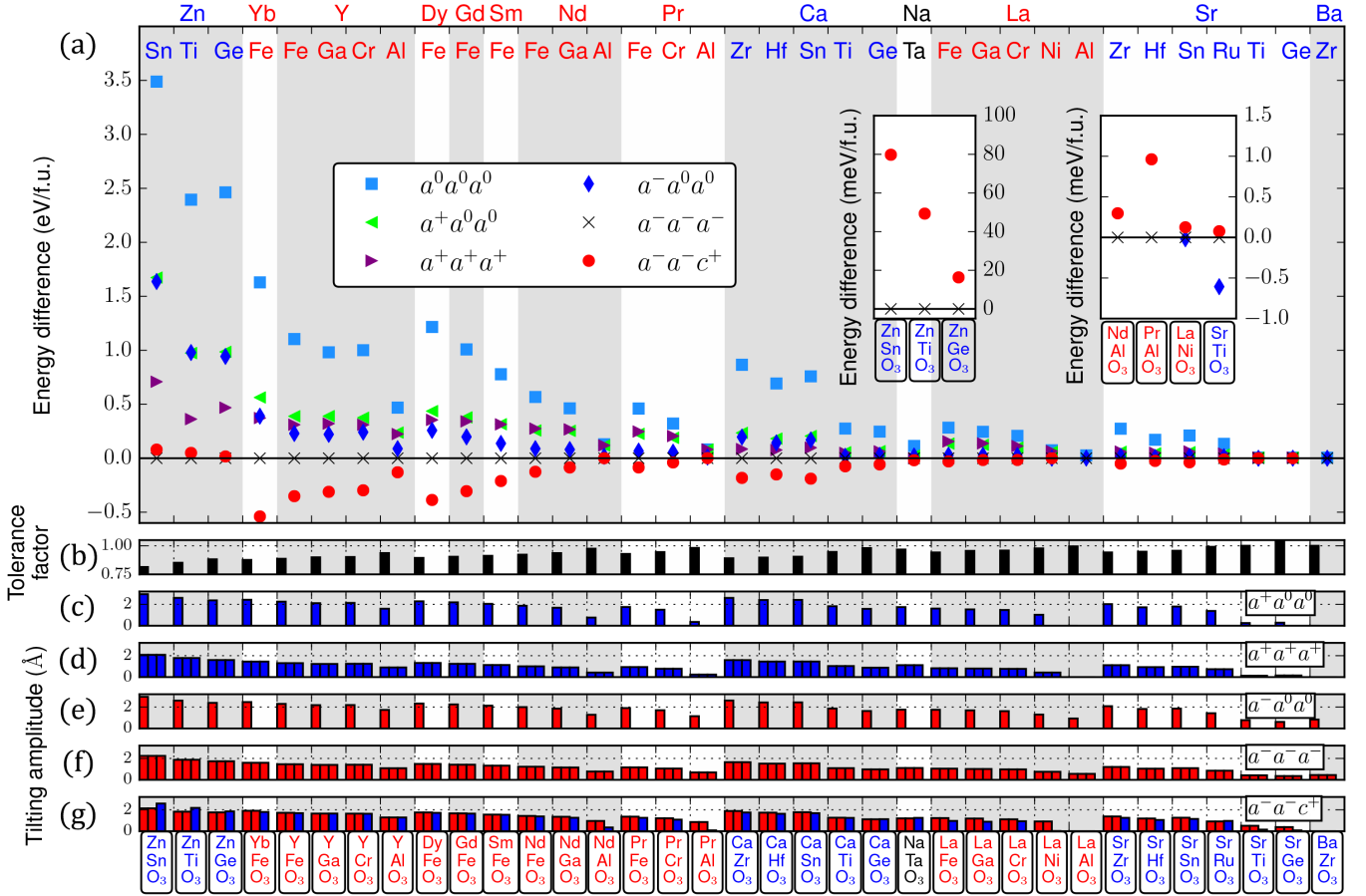}
\caption{Summary of our first-principles results. Panel~(a) shows the
  energies of the different tilt phases considered, given in eV per
  formula unit (f.u.). We take the result for the $a^{-}a^{-}a^{-}$
  structure as zero of energy. The insets display zooms of the results
  for the materials that do not present the $a^{-}a^{-}c^{+}$ ground
  state (energies in meV/f.u.). The case of SrGeO$_{3}$ is not visible
  even in the inset; for this compound we obtain an $a^{-}b^{0}b^{0}$
  ground state that is only 0.025~meV/f.u. below the $a^{-}a^{-}a^{-}$
  structure. Panel~(b) shows the tolerance factor $t$ of the
  considered compounds. Note that the compounds are ordered from left
  to right as follows: We place together all the materials that share
  the same {\sl A} cation, and the ionic radius of {\sl A} grows as we
  move to the right. Compounds sharing the same {\sl A} cation are
  ordered so that the ionic radius of the {\sl B} cation decreases as
  we move to the right. All in all, the tolerance factor roughly grows
  when we move from left to right. Panels(c)--(g) show the antiphase
  and in-phase rotation components (\AA) corresponding to the relaxed
  structures. The chemical formulas are given following a color code,
  red corresponding to compounds with {\sl A}$^{3+}${\sl
    B}$^{3+}$O$_{3}^{2-}$ nominal ionizations, blue to {\sl
    A}$^{2+}${\sl B}$^{4+}$O$_{3}^{2-}$, and black to {\sl
    A}$^{1+}${\sl B}$^{5+}$O$_{3}^{2-}$.}
\label{fig:rawresults}
\end{figure*}

\section{Results and Discussion}
\label{sec:results}

In the following we present our results and discuss their
implications, paying especial attention to the reasons why the $Pbnm$
structure is the ground state of most perovskite oxides.

\subsection{Raw first-principles results}

Figure~\ref{fig:rawresults} summarizes our raw results, from which
many conclusions can be directly drawn. Let us go over them in an
orderly manner, and in the next Section we will see how each of them
is reflected in (and explained by) the parameters of our energy
function.

First, the energy difference between the reference cubic structure and
the lowest-energy (ground state) configuration varies strongly with
the tolerance factor, from about 3.5~eV/f.u. for ZnSnO$_{3}$ to less
than 5~meV/f.u. for BaZrO$_{3}$. Correspondingly, there is a decrease
of the AFD distortion amplitudes for increasing $t$, although not as
drastic; for example, we have 2.3~\AA\ for each of the three
components of ${\bf r}$ in the $a^{-}a^{-}a^{-}$ state of ZnSnO$_{3}$,
while we obtain $r_{x} \approx 0.8$~\AA\ in the $a^{-}b^{0}b^{0}$
state of BaZrO$_{3}$. (These distortion amplitudes may seem
unrealistically large; this is a consequence of our choice for the
normalization of the associated symmetry-adapted vectors, described
above.)  Note that the mentioned energy gap -- and, to a lesser
extent, the distortion amplitudes -- are expected to correlate with
the temperature at which the cubic phase would transform into the
low-symmetry structure, bigger energy differences -- or bigger
distortions -- corresponding to higher-temperature
transitions.\cite{abrahams68,wojdel14b} Our results are consistent
with the experimental observations in this regard. For example,
intermediate-$t$ compound CaTiO$_{3}$ remains tilted up to about
1500~K,\cite{redfern96} while LaFeO$_{3}$ is estimated to become cubic
above 2000~K (provided it does not melt first).\cite{selbach12} In
contrast, the cubic phase of our materials with $t \approx 1$ is
stable at rather low temperatures, e.g., down to 110~K in SrTiO$_{3}$
(Refs.~\onlinecite{lytle64,fleury68}) and down to essentially 0~K in
BaZrO$_{3}$, for which the symmetry-breaking distortions are probably
suppressed by quantum fluctuations.\cite{akbarzadeh05} Finally, let us
note that, as a consequence of the weaker AFD instabilities, the
energy differences between tilt polymorphs become very small for
large-$t$ compounds; generally, this should result in a greater
structural tunability (e.g., by means of epitaxial strain imposed on
thin films) in these materials.

Second, for all the investigated compounds, the antiphase O$_{6}$
rotations render stronger structural instabilities than their in-phase
counterparts. Both instability types behave in a much correlated
manner, becoming simultaneously stronger, or weaker, as a function of
$t$. Interestingly, for $t \lesssim 1$, we find compounds in which the
${\bf r}$-tilts are still a (weak) instability of the cubic structure
while the ${\bf m}$-tilts are not. Examples of this are LaAlO$_{3}$
and BaZrO$_{3}$, for which we find it impossible to relax
$a^{+}b^{0}b^{0}$ or $a^{+}a^{+}a^{+}$ phases. (In those simulations
the compounds relax back to the cubic reference structure; the
corresponding results are missing in Fig.~\ref{fig:rawresults}.) It is
not our task here to investigate the atomistic reasons for the general
-- albeit slight -- prevalence of antiphase tilt patterns over
in-phase ones; let us note, though, that a discussion of this matter
can be found in Ref.~\onlinecite{gu17}.

Third, the ${\cal O}$ phase $a^{-}a^{-}c^{+}$ is not obtained as a
distinct solution for two materials, namely, LaAlO$_{3}$ and
BaZrO$_{3}$. In such cases, during the structural relaxation of the
${\cal O}$ phase -- for which we use a typical $a^{-}a^{-}c^{+}$
configuration as starting point -- we observe a progressive decrease
of the in-phase $m_{z}$ rotation, until the $a^{-}a^{-}c^{0}$ solution
is obtained as final result. Note that these are exactly the same two
compounds for which we cannot stabilize structures with only in-phase
tilts.

Fourth, the ${\cal O}$ phase is the most stable structure (i.e., the
ground state) of the majority of studied materials. On one hand, while
our batch of crystals is obviously a limited one, this observation is
consistent with reality. Indeed, it is well-known that the ${\cal O}$
phase is dominant among perovskite oxides and, in particular, the
number of compounds displaying tilt structures other than
$a^{-}a^{-}c^{+}$ is comparatively small.\cite{lufaso01} On the other
hand, if we take into account the points made above, this is a
somewhat surprising result. Indeed, our calculations show the
preeminence of antiphase tilts over their in-phase counterparts, which
suggests that purely antiphase patterns should be dominant over purely
in-phase ones (as is indeed the case) and over antiphase/in-phase
combinations as well (obviously not the case). Then, to explain why
the ${\cal O}$ phase is generally preferred, it would be most natural
to imagine some sort of cooperative interaction between antiphase and
in-phase rotations, which would drive their simultaneous
occurrence. However, our results clearly suggest that, for the $Pbnm$
state to exist, the in-phase tilts must be a native instability of the
cubic structure, which seems at odds with the cooperation
hypothesis. Further, if the ${\bf r}$ and ${\bf m}$ rotations were to
cooperate, we would expect to see an enhancement of their amplitudes
when they appear combined in the ${\cal O}$ phase; however, this is
not observed in our results. Hence, the dominance of the $Pbnm$ ground
state is a surprise that we cannot explain from the results presented
thus far.

Finally, if we look at the compounds that present lowest-energy
structures other than $Pbnm$, they clearly belong to two different
classes. On one hand, we have a group of large-$t$ materials in which
the in-phase tilts are either a weak instability of the cubic phase
(NdAlO$_{3}$, PrAlO$_{3}$, LaNiO$_{3}$, SrTiO$_{3}$, and SrGeO$_{3}$)
or not unstable at all (LaAlO$_{3}$ and BaZrO$_{3}$). This clearly
suggests that relatively strong in-phase tilts are necessary to obtain
a $Pbnm$ ground state. On the other hand, we have small-$t$ compounds
(ZnSnO$_{3}$, ZnTiO$_{3}$, and ZnGeO$_{3}$) for which all considered
AFD distortions are very strong instabilities of the cubic
phase. However, for such materials the $a^{-}a^{-}a^{-}$ state lies
lower in energy than $a^{-}a^{-}c^{+}$. Naturally, the reasons behind
these results in the small-$t$ limit must be quite different from
those relevant to large-$t$ materials like SrTiO$_{3}$ or
BaZrO$_{3}$. In particular, it is interesting to note that, as
mentioned above, the considered small-$t$ Zn-based compounds are not
perovskites in reality; thus, one may wonder whether their
predilection for other crystalline lattices (LiNbO$_{3}$-like and
ilmenite) may be related to their preference for $a^{-}a^{-}a^{-}$
over $a^{-}a^{-}c^{+}$.

\begin{table*}[htbp]
  \renewcommand\thetable{I}
  \centering
  \caption{Calculated parameters for the energy $E({\bf r},{\bf
      m},\{\eta_a\})$ of Eq.~(\ref{eq:energy}). The harmonic constants
    $\kappa_r$ and $\kappa_m$ are given in eV/\AA$^2$. The anharmonic
    couplings $\alpha_r$, $\alpha_m$, $\gamma_r$, $\gamma_m$,
    $\alpha_{int}$, and $\beta_{int}$ are given in eV/\AA$^4$. The
    elastic constants $C_{11}$, $C_{12}$ and $C_{44}$ are given in
    eV. The strain-phonon couplings $B_{1rxx}$, $B_{1ryy}$,
    $B_{4ryz}$, $B_{1mxx}$, and $B_{1myy}$ are given in
    eV/\AA$^2$. The lattice constant of the reference
      cubic cell ($a$) is given in \AA.} See
      Section~\ref{sec:methods} for more details on the units and
      normalization used. For compounds in which the in-phase tilts
    are not instabilities of the cubic phase ($\kappa_{m} > 0$), the
    corresponding anharmonic and strain-phonon couplings are not
    given.
    \begin{ruledtabular}
    \begin{tabular*}{\textheight}{lrrrrrrrrrrrrrrr}
        & \multicolumn{1}{c}{$\kappa_r$} & \multicolumn{1}{c}{$\alpha_r$} & \multicolumn{1}{c}{$\gamma_r$} & \multicolumn{1}{c}{$\alpha_{int}$} & \multicolumn{1}{c}{$B_{1rxx}$} & \multicolumn{1}{c}{$B_{1ryy}$} & \multicolumn{1}{c}{$B_{4ryz}$} & \multicolumn{1}{c}{$C_{12}$} & \multirow{2}{*}{$a$}\\
        & \multicolumn{1}{c}{$\kappa_m$} & \multicolumn{1}{c}{$\alpha_m$} & \multicolumn{1}{c}{$\gamma_m$} &  \multicolumn{1}{c}{$\beta_{int}$} & \multicolumn{1}{c}{$B_{1mxx}$} & \multicolumn{1}{c}{$B_{1myy}$} & \multicolumn{1}{c}{$C_{11}$} & \multicolumn{1}{c}{$C_{44}$} & \\
    \hline 
    \multirow{2}{*}{\chemform{ZnSnO_3}} & $-$3.4368 & 0.2116 & $-$0.2853 & 0.2085 & 5.1551 & 2.5882 & $-$0.6064& 332.02 & \multirow{2}{*}{3.988}\\
                                        & $-$3.3607 & 0.2108 & $-$0.2040 & 0.8413 & 4.9309 & 3.7545 & 962.49 & 127.55 & \\
    \multirow{2}{*}{\chemform{ZnTiO_3}} & $-$3.3731 & 0.2665 & $-$0.3163 & 0.2336 & 5.6122 & 2.6985 &$-$0.9868 & 357.63 & \multirow{2}{*}{3.800} \\
                                        & $-$3.3629 & 0.2696 & $-$0.2278 & 1.2845 & 5.5464 & 4.2004 & 913.22 & 149.10 & \\  
    \multirow{2}{*}{\chemform{ZnGeO_3}} & $-$4.2563 & 0.3821 & $-$0.4359 & 0.3051 & 4.0894 & 1.7946 &$-$1.1295 & 393.32 & \multirow{2}{*}{3.679} \\
                                        & $-$4.2129 & 0.3850 & $-$0.2940 & 1.7780 & 3.7574 & 2.7450 & 800.93 & 243.53 & \\
    \multirow{2}{*}{\chemform{YbFeO_3}} & $-$3.2666 & 0.2690 & $-$0.1914 & 0.0875 & 1.2414 & 0.1461 & 0.8969 & 311.61 & \multirow{2}{*}{3.795} \\
                                        & $-$2.9343 & 0.2493 & $-$0.1099 & 1.7128 &$-$0.0025 &$-$0.1914 & 999.77 & 189.83 & \\
    \multirow{2}{*}{\chemform{YFeO_3}}  & $-$2.6636 & 0.2525 & $-$0.1527 & 0.0933 & 0.1725 & 0.0676 & 1.2268 & 323.35 & \multirow{2}{*}{3.827} \\
                                        & $-$2.2967 & 0.2279 & $-$0.0634 & 1.6222 &$-$1.2182 &$-$0.3470 & 982.19 & 216.52 & \\
    \multirow{2}{*}{\chemform{YGaO_3}}  & $-$2.5675 & 0.2702 & $-$0.1763 & 0.0824 &$-$0.7934 &$-$0.2877 & 1.6840 & 334.86 & \multirow{2}{*}{3.793} \\
                                        & $-$2.1638 & 0.2466 & $-$0.0742 & 1.7953 &$-$2.4215 &$-$0.8336 & 998.33 & 280.62 & \\
    \multirow{2}{*}{\chemform{YCrO_3}}  & $-$2.5499 & 0.2689 & $-$0.1936 & 0.1002 &$-$1.5850 & 0.7523 & 1.4620 & 226.79 & \multirow{2}{*}{3.776} \\
                                        & $-$2.2363 & 0.2505 & $-$0.0728 & 1.7821 &$-$2.8801 & 0.0463 &1147.72 & 195.55 & \\
    \multirow{2}{*}{\chemform{YAlO_3}}  & $-$2.0325 & 0.3427 & $-$0.1882 & 0.1006 &$-$2.8200 &$-$1.0237 & 2.3993 & 311.32 & \multirow{2}{*}{3.683} \\
                                        & $-$1.4610 & 0.2949 & $-$0.0546 & 2.2546 &$-$2.4582 &$-$2.5669 &1011.53 & 386.22 & \\
    \multirow{2}{*}{\chemform{DyFeO_3}} & $-$2.8305 & 0.2606 & $-$0.1611 & 0.0918 & 0.2004 & 0.0485 & 1.2369 & 318.60 & \multirow{2}{*}{3.814} \\
                                        & $-$2.4361 & 0.2352 & $-$0.0615 & 1.6331 &$-$1.3442 &$-$0.4068 &1003.26 & 213.18 & \\
    \multirow{2}{*}{\chemform{GdFeO_3}} & $-$2.5751 & 0.2549 & $-$0.1445 & 0.0954 &$-$0.4067 &$-$0.0846 & 1.4513 & 321.96 & \multirow{2}{*}{3.825} \\
                                        & $-$2.1526 & 0.2283 & $-$0.0335 & 1.6022 &$-$2.1072 &$-$0.4935 &1001.97 & 226.86 & \\
    \multirow{2}{*}{\chemform{SmFeO_3}} & $-$2.2626 & 0.2497 & $-$0.1267 & 0.1023 &$-$1.2798 &$-$0.2075 & 1.7157 & 327.46 & \multirow{2}{*}{3.842} \\
                                        & $-$1.8100 & 0.2221 & $-$0.0019 & 1.5536 &$-$3.1754 &$-$0.7478 &1000.37 & 243.24 & \\
    \multirow{2}{*}{\chemform{NdFeO_3}} & $-$1.9300 & 0.2445 & $-$0.1086 & 0.1135 &$-$2.0486 &$-$0.4774 & 2.0031 & 336.17 & \multirow{2}{*}{3.861} \\
                                        & $-$1.4556 & 0.2192 &  0.0296 & 1.4780 &$-$4.2207 &$-$0.8632 & 998.83 & 261.69 & \\
    \multirow{2}{*}{\chemform{NdGaO_3}} & $-$1.7823 & 0.2642 & $-$0.1323 & 0.1039 &$-$3.1943 &$-$0.7815 & 2.5818 & 349.72 & \multirow{2}{*}{3.835} \\
                                        & $-$1.2038 & 0.2335 &  0.0149 & 1.6207 &$-$6.1151 &$-$1.2102 &1010.16 & 323.86 & \\
    \multirow{2}{*}{\chemform{NdAlO_3}} & $-$1.0916 & 0.3501 & $-$0.1465 & 0.1328 &$-$5.9629 &$-$1.9274 & 3.4533 & 339.94 & \multirow{2}{*}{3.733} \\
                                        & $-$0.3287 & 0.3395 &  0.0224 & 2.0994 &$-$10.5517&$-$2.2041 &1019.30 & 435.80 & \\
    \multirow{2}{*}{\chemform{PrFeO_3}} & $-$1.7410 & 0.2427 & $-$0.0974 & 0.1219 &$-$2.5728 &$-$0.8180 & 2.1678 & 341.47 & \multirow{2}{*}{3.873} \\
                                        & $-$1.2481 & 0.2175 &  0.0496 & 1.4386 &$-$5.0120 &$-$1.1222 & 998.60 & 272.66 & \\
    \multirow{2}{*}{\chemform{PrCrO_3}} & $-$1.4666 & 0.2653 & $-$0.1461 & 0.1342 &$-$5.2617 &$-$0.2739 & 2.7027 & 269.95 & \multirow{2}{*}{3.831} \\
                                        & $-$0.9582 & 0.2440 &  0.0316 & 1.5880 &$-$8.1167 &$-$0.3844 &1212.03 & 269.25 & \\
    \multirow{2}{*}{\chemform{PrAlO_3}} & $-$0.8632 & 0.3476 & $-$0.1390 & 0.0389 &$-$6.4502 &$-$2.0775 & 3.5909 & 346.35 & \multirow{2}{*}{3.750} \\
                                        & $-$0.0663 & 0.3276 &  0.0300 & 2.3272 &$-$12.0365&$-$1.6655 &1011.89 & 443.97 & \\
    \multirow{2}{*}{\chemform{CaZrO_3}} & $-$1.6336 & 0.1440 & $-$0.0946 & 0.1165 & 1.9993 & 5.5824 &$-$0.0627 & 250.60 & \multirow{2}{*}{4.123} \\
                                        & $-$1.5521 & 0.1407 & $-$0.0840 & 0.7943 & 1.6820 & 5.8031 &1265.58 & 213.56 & \\
    \multirow{2}{*}{\chemform{CaHfO_3}} & $-$1.5474 & 0.1583 & $-$0.0994 & 0.1264 & 1.6587 & 6.0322 & $-$0.021 & 248.32 & \multirow{2}{*}{4.064} \\
                                        & $-$1.4616 & 0.1547 & $-$0.0891 & 0.8926 & 1.2879 & 6.2055 &1323.30 & 240.74 & \\
    \multirow{2}{*}{\chemform{CaSnO_3}} & $-$1.6244 & 0.1516 & $-$0.1082 & 0.0956 & 0.3449 & 3.9340 & 0.4804 & 284.04 & \multirow{2}{*}{4.028} \\
                                        & $-$1.5623 & 0.1498 & $-$0.0865 & 0.9434 &$-$0.0528 & 3.9019 &1074.87 & 238.27 & \\
    \multirow{2}{*}{\chemform{CaTiO_3}} & $-$1.2063 & 0.1985 & $-$0.0658 & 0.1376 & 0.0555 & 5.1580 & 0.7306 & 294.12 & \multirow{2}{*}{3.846} \\
                                        & $-$1.0870 & 0.1917 & $-$0.0543 & 1.2283 &$-$0.5534 & 5.2159 &1062.00 & 281.25 & \\
    \multirow{2}{*}{\chemform{CaGeO_3}} & $-$1.3360 & 0.2840 & $-$0.1651 & 0.1473 &$-$4.2548 & 2.9271 & 2.1924 & 313.88 & \multirow{2}{*}{3.739} \\
                                        & $-$1.1928 & 0.2802 & $-$0.1128 & 1.8836 &$-$5.3103 & 2.5564 & 900.20 & 361.11 & \\
    \multirow{2}{*}{\chemform{NaTaO_3}} & $-$0.4960 & 0.1530 & $-$0.1056 & 0.1493 & 2.0430 &10.9583 &$-$0.8661 & 227.63 & \multirow{2}{*}{3.946} \\
                                        & $-$0.4739 & 0.1527 & $-$0.1173 & 0.8557 & 2.0712 &11.2295 &1482.53 & 245.99 & \\
    \multirow{2}{*}{\chemform{LaFeO_3}} & $-$1.3433 & 0.2255 & $-$0.0650 & 0.1639 &$-$3.0822 &$-$0.9321 & 2.0223 & 353.27 & \multirow{2}{*}{3.901} \\
                                        & $-$0.9311 & 0.1962 &  0.0613 & 1.1851 &$-$5.1743 &$-$1.0036 & 983.43 & 283.46 & \\
    \multirow{2}{*}{\chemform{LaGaO_3}} & $-$1.2211 & 0.2205 & $-$0.0738 & 0.1824 &$-$3.8291 &$-$1.2673 & 2.0206 & 365.85 & \multirow{2}{*}{3.864} \\
                                        & $-$0.8453 & 0.2008 &  0.0405 & 1.1222 &$-$5.8161 &$-$1.3636 & 986.80 & 329.12 & \\
    \multirow{2}{*}{\chemform{LaCrO_3}} & $-$1.1160 & 0.2258 & $-$0.0937 & 0.1821 &$-$5.4157 &$-$0.5899 & 2.1836 & 277.65 & \multirow{2}{*}{3.848} \\
                                        & $-$0.8144 & 0.2134 &  0.0327 & 1.1975 &$-$7.2818 &$-$0.3057 &1192.41 & 270.53 & \\
    \multirow{2}{*}{\chemform{LaNiO_3}} & $-$0.7139 & 0.2965 & $-$0.1731 & 0.0779 &$-$11.8529& 0.0002 & 4.3395 & 362.23 & \multirow{2}{*}{3.769} \\
                                        & $-$0.2516 & 0.2542 & $-$0.0278 & 1.9808 &$-$14.6453& 0.6517 &1041.72 & 296.06 & \\
    \multirow{2}{*}{\chemform{LaAlO_3}} & $-$0.4669 & 0.2925 & $-$0.0623 & ------ &$-$6.9410 &$-$1.9893 & 2.3526 & 357.29 & \multirow{2}{*}{3.771} \\
                                        &  0.1136 & ------ &  ------ & ------ & ------ & ------ & 989.05 & 431.82 & \\
    \multirow{2}{*}{\chemform{SrZrO_3}} & $-$0.9691 & 0.1363 & $-$0.0517 & 0.1242 & 0.0948 & 5.5680 & 0.5546 & 265.56 & \multirow{2}{*}{4.156} \\
                                        & $-$0.8696 & 0.1327 & $-$0.0330 & 0.7631 &$-$0.5006 & 5.5736 &1254.67 & 258.88 & \\
    \end{tabular*}%
    \end{ruledtabular}
  \label{tab:bare-parameters}%
\end{table*}%

\begin{table*}[htbp]
  \renewcommand\thetable{I}
  \centering
  \caption{(Continued.)}
    \begin{ruledtabular}
    \begin{tabular*}{\textheight}{lrrrrrrrrrrrrrrr}
        & \multicolumn{1}{c}{$\kappa_r$} & \multicolumn{1}{c}{$\alpha_r$} & \multicolumn{1}{c}{$\gamma_r$} & \multicolumn{1}{c}{$\alpha_{int}$} & \multicolumn{1}{c}{$B_{1rxx}$} & \multicolumn{1}{c}{$B_{1ryy}$} & \multicolumn{1}{c}{$B_{4ryz}$} & \multicolumn{1}{c}{$C_{12}$} & \multirow{2}{*}{$a$}\\
        & \multicolumn{1}{c}{$\kappa_m$} & \multicolumn{1}{c}{$\alpha_m$} & \multicolumn{1}{c}{$\gamma_m$} &  \multicolumn{1}{c}{$\beta_{int}$} & \multicolumn{1}{c}{$B_{1mxx}$} & \multicolumn{1}{c}{$B_{1myy}$} & \multicolumn{1}{c}{$C_{11}$} & \multicolumn{1}{c}{$C_{44}$} & \\
    \hline 
    \multirow{2}{*}{\chemform{SrHfO_3}} & $-$0.8050 & 0.1487 & $-$0.0538 & 0.1411 &$-$0.4475 & 5.9698 & 0.7210 & 266.54 & \multirow{2}{*}{4.098} \\
                                        & $-$0.6982 & 0.1471 & $-$0.0335 & 0.8377 &$-$1.2223 & 5.8840 &1306.46 & 290.66 & \\
    \multirow{2}{*}{\chemform{SrSnO_3}} & $-$0.9135 & 0.1511 & $-$0.0722 & 0.1008 &$-$2.3470 & 3.5175 & 1.3626 & 288.47 & \multirow{2}{*}{4.067} \\
                                        & $-$0.7916 & 0.1443 & $-$0.0350 & 0.9393 &$-$3.1308 & 3.2244 &1068.14 & 285.23 & \\
    \multirow{2}{*}{\chemform{SrRuO_3}} & $-$0.9559 & 0.2509 & $-$0.0953 & 0.2353 &$-$3.0008 & 2.0584 & 2.4588 & 334.02 & \multirow{2}{*}{3.908} \\
                                        & $-$0.8691 & 0.2466 &  0.0428 & 1.3700 &$-$4.2993 & 0.9725 &1025.70 & 218.88 & \\
    \multirow{2}{*}{\chemform{SrTiO_3}} & $-$0.1882 & 0.1865 & $-$0.0242 & 0.1393 &$-$3.1262 & 4.3480 & 1.8653 & 312.45 & \multirow{2}{*}{3.899} \\
                                        & $-$0.0215 & 0.1842 &  0.0614 & 1.3014 &$-$4.5659 & 4.8710 &1054.96 & 337.56 & \\
    \multirow{2}{*}{\chemform{SrGeO_3}} & $-$0.2585 & 0.4252 & $-$0.1934 & 0.0827 &$-$9.6994 & 1.4051 & 4.2523 & 299.63 & \multirow{2}{*}{3.807} \\
                                        & $-$0.0928 & 0.6012 &  0.2004 & 3.6661 &$-$12.8138& 0.6900 & 858.38 & 415.25 & \\
    \multirow{2}{*}{\chemform{BaZrO_3}} & $-$0.1135 & 0.1013 & $-$0.0288 & ------ &$-$2.0365 & 4.2014 & 1.2545 & 303.73 & \multirow{2}{*}{4.210} \\
                                        &  0.1226 & ------ &  ------ & ------ & ------ & ------ &1224.42 & 315.85 & \\
    \end{tabular*}%
    \end{ruledtabular}
\end{table*}%

\begin{table*}[htbp]
  \renewcommand\thetable{II}
  \centering
  \caption{Calculated parameters for the energy $\bar{E}({\bf r},{\bf
      m})$ of Eq.~(\ref{eq:renorm-energy}). The harmonic constants
    $\kappa_r$ and $\kappa_m$ are given in eV/\AA$^2$. The anharmonic
    couplings $\bar{\alpha}_r$, $\bar{\alpha}_m$, $\bar{\gamma}_r$,
    $\bar{\gamma}_m$, $\bar{\alpha}_{int}$, and $\bar{\beta}_{int}$
    are in eV/\AA$^4$. The 6th-order correction $\bar{\gamma}_{\rm
      int}$ (introduced in Section~\ref{sec:sixth-order}) is in
    eV/\AA$^{6}$. See Section~\ref{sec:methods} for more
      details on the units and normalization used. For compounds in
    which the in-phase tilts are not instabilities of the cubic phase
    ($\kappa_{m} > 0$), the corresponding anharmonic couplings are not
    given. For compounds in which the $a^{+}b^{-}a^{+}$ phase is not a
    local energy minimum, the $\bar{\gamma}_{\rm int}$ coupling is not
    given.}
    \begin{ruledtabular}
    \begin{tabular*}{\textheight}{lrrrrrrrrr}
          & \multicolumn{1}{c}{$\kappa_r$} & \multicolumn{1}{c}{$\bar{\alpha}_r$} & \multicolumn{1}{c}{$\bar{\gamma}_r$} & \multicolumn{1}{c}{$\kappa_m$} & \multicolumn{1}{c}{$\bar{\alpha}_m$} & \multicolumn{1}{c}{$\bar{\gamma}_m$} & \multicolumn{1}{c}{$\bar{\alpha}_{int}$} & \multicolumn{1}{c}{$\bar{\beta}_{int}$} &  \multicolumn{1}{c}{$\bar{\gamma}_{\rm int}$}\\
    \hline 
   \chemform{ZnSnO_3}	 & -3.3806 & 0.1929 & -0.2745 & -3.3401 & 0.1920 & -0.2001 & 0.1254 & 1.0073 & 0.0066 \\
   \chemform{ZnTiO_3}	 & -3.3355 & 0.2456 & -0.3045 & -3.3593 & 0.2481 & -0.2241 & 0.1970 & 1.2989 & 0.0131 \\
   \chemform{ZnGeO_3}	 & -4.2493 & 0.3712 & -0.4268 & -4.2089 & 0.3746 & -0.2912 & 0.2873 & 1.7852 & 0.0489 \\
   \chemform{YbFeO_3}	 & -3.2585 & 0.2677 & -0.1936 & -2.9015 & 0.2468 & -0.1180 & 0.0924 & 1.6857 & 0.1643 \\
   \chemform{YFeO_3}	 & -2.6538 & 0.2516 & -0.1580 & -2.2627 & 0.2237 & -0.0719 & 0.1011 & 1.5832 & 0.2172 \\
   \chemform{YGaO_3}	 & -2.5565 & 0.2688 & -0.1832 & -2.1307 & 0.2400 & -0.0812 & 0.0926 & 1.7445 & 0.2626 \\
   \chemform{YCrO_3}	 & -2.5369 & 0.2648 & -0.1934 & -2.1990 & 0.2415 & -0.0724 & 0.1122 & 1.7274 & 0.2086 \\
   \chemform{YAlO_3}	 & -2.0270 & 0.3379 & -0.1934 & -1.4619 & 0.2893 & -0.0545 & 0.1032 & 2.2250 & 0.8270 \\
   \chemform{DyFeO_3}	 & -2.8213 & 0.2597 & -0.1663 & -2.3976 & 0.2306 & -0.0710 & 0.0996 & 1.5926 & 0.2196 \\
   \chemform{GdFeO_3}	 & -2.5649 & 0.2538 & -0.1510 & -2.1130 & 0.2215 & -0.0418 & 0.1048 & 1.5532 & 0.2706 \\
   \chemform{SmFeO_3}	 & -2.2550 & 0.2479 & -0.1324 & -1.7730 & 0.2120 & -0.0073 & 0.1122 & 1.4981 & 0.3831 \\
   \chemform{NdFeO_3}	 & -1.9220 & 0.2412 & -0.1143 & -1.4237 & 0.2044 &  0.0319 & 0.1228 & 1.4180 & 0.6398 \\
   \chemform{NdGaO_3}	 & -1.7735 & 0.2575 & -0.1356 & -1.1787 & 0.2084 &  0.0385 & 0.1131 & 1.5506 & 0.9400 \\
   \chemform{NdAlO_3}	 & -1.0897 & 0.3320 & -0.1375 & -0.3255 & 0.2793 &  0.1199 & 0.1203 & 2.0454 & ------ \\
   \chemform{PrFeO_3}	 & -1.7331 & 0.2383 & -0.1037 & -1.2218 & 0.1993 &  0.0587 & 0.1290 & 1.3800 & 1.0181 \\
   \chemform{PrCrO_3}	 & -1.4609 & 0.2517 & -0.1338 & -0.9464 & 0.2113 &  0.0878 & 0.1476 & 1.5077 & 1.0809 \\
   \chemform{PrAlO_3}	 & -0.8618 & 0.3264 & -0.1259 & -0.0666 & 0.2534 &  0.1887 & 0.0345 & 2.2456 & ------ \\
   \chemform{CaZrO_3}	 & -1.5673 & 0.1146 & -0.0797 & -1.4956 & 0.1107 & -0.0664 & 0.0876 & 0.7610 & 0.0320 \\
   \chemform{CaHfO_3}	 & -1.4884 & 0.1259 & -0.0797 & -1.4151 & 0.1225 & -0.0660 & 0.0992 & 0.8508 & 0.0456 \\
   \chemform{CaSnO_3}	 & -1.5851 & 0.1333 & -0.0909 & -1.5194 & 0.1304 & -0.0669 & 0.0900 & 0.9035 & 0.0566 \\
   \chemform{CaTiO_3}	 & -1.1794 & 0.1692 & -0.0337 & -1.0602 & 0.1598 & -0.0133 & 0.1370 & 1.1456 & 0.3083 \\
   \chemform{CaGeO_3}	 & -1.3153 & 0.2472 & -0.0838 & -1.1631 & 0.2342 & -0.0133 & 0.1885 & 1.7293 & 0.4665 \\
   \chemform{NaTaO_3}	 & -0.4816 & 0.0780 & -0.0418 & -0.4708 & 0.0777 & -0.0499 & 0.0693 & 0.7900 & 0.0364 \\
   \chemform{LaFeO_3}	 & -1.3379 & 0.2197 & -0.0667 & -0.9151 & 0.1777 &  0.0782 & 0.1649 & 1.1464 & 2.3807 \\
   \chemform{LaGaO_3}	 & -1.2144 & 0.2117 & -0.0720 & -0.8300 & 0.1786 &  0.0618 & 0.1794 & 1.0847 & 1.9170 \\
   \chemform{LaCrO_3}	 & -1.1087 & 0.2113 & -0.0790 & -0.8049 & 0.1863 &  0.0803 & 0.1884 & 1.1390 & 0.8844 \\
   \chemform{LaNiO_3}	 & -0.7014 & 0.2076 &  0.0001 & -0.2457 & 0.1172 &  0.3008 & 0.1599 & 1.6352 & ------ \\
   \chemform{LaAlO_3}	 & -0.4662 & 0.2672 & -0.0303 &  0.1136 & ------ &  ------ & ------ & ------ & ------ \\
   \chemform{SrZrO_3}	 & -0.9428 & 0.1087 & -0.0230 & -0.8419 & 0.1033 &  0.0012 & 0.1166 & 0.6996 & 0.1392 \\
   \chemform{SrHfO_3}	 & -0.7871 & 0.1184 & -0.0161 & -0.6790 & 0.1147 &  0.0115 & 0.1377 & 0.7645 & 0.2838 \\
   \chemform{SrSnO_3}	 & -0.8969 & 0.1297 & -0.0317 & -0.7701 & 0.1198 &  0.0123 & 0.1183 & 0.8592 & 0.2798 \\
   \chemform{SrRuO_3}	 & -0.9374 & 0.2306 & -0.0741 & -0.8382 & 0.2202 &  0.0633 & 0.2549 & 1.2706 & 0.4282 \\
   \chemform{SrTiO_3}	 & -0.1857 & 0.1544 &  0.0441 & -0.0215 & 0.1818 &  0.0606 & 0.1559 & 1.2345 & ------ \\
   \chemform{SrGeO_3}	 & -0.2552 & 0.3382 &  0.0040 & -0.0928 & 0.5876 &  0.1956 & 0.0887 & 3.5973 & ------ \\
   \chemform{BaZrO_3}	 & -0.1101 & 0.0789 &  0.0079 &  0.1226 & ------ &  ------ & ------ & ------ & ------ \\
    \end{tabular*}%
    \end{ruledtabular}
    \label{tab:strain-renorm-parameters}%
\end{table*}%

\subsection{Modeling the relevant potential energy surface}

\begin{figure*}
\includegraphics[width=1.00\linewidth]{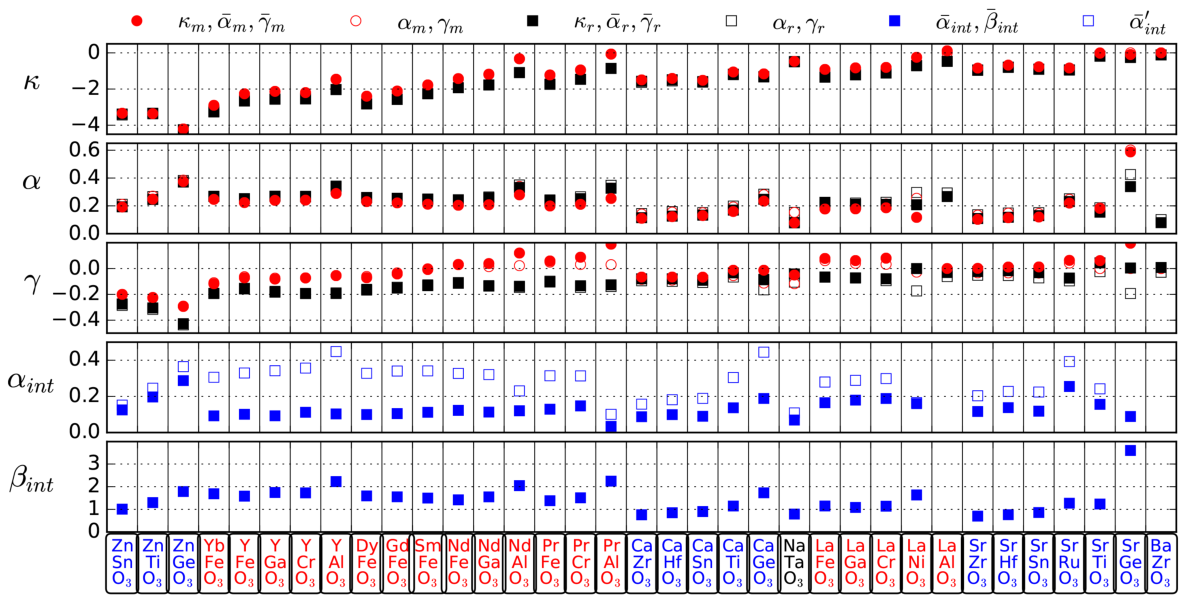}
\caption{Computed PES parameters. We use: solid black squares for
  $\kappa_{r}$, $\bar{\alpha}_{r}$, and $\bar{\gamma}_{r}$; solid red
  circles for $\kappa_{m}$, $\bar{\alpha}_{m}$, and
  $\bar{\gamma}_{m}$; solid blue squares for $\bar{\alpha}_{\rm int}$
  and $\bar{\beta}_{\rm int}$; open black squares for $\alpha_{r}$ and
  $\gamma_{r}$; open red circles for $\alpha_{m}$ and $\gamma_{m}$;
  open blue squares for $\bar{\alpha}'_{\rm int}$. The color code for
  the chemical formulas is as in Fig.~\ref{fig:rawresults}.}
\label{fig:parameters}
\end{figure*}

Next, we use the results described above to fit the parameters
defining the relevant PES, following the guidelines given in
Section~\ref{sec:methods}. Table~\ref{tab:bare-parameters} shows the
results obtained for the parameters entering the energy of
Eq.~(\ref{eq:energy}), where strains are explicitly considered. In
contrast, in Table~\ref{tab:strain-renorm-parameters} we present the
results obtained for the parameters that implicitly capture the strain
relaxations that follow the primary orders ${\bf r}$ and ${\bf m}$,
corresponding to Eq.~(\ref{eq:renorm-energy}). Finally,
Fig.~\ref{fig:parameters} displays the key couplings in a way that
makes it easier to appreciate trends as a function of the tolerance
factor. For simplicity, in this Section we focus on the
strain-renormalized results to discuss the main features of the
PES. The computed parameters reflect and explain the conclusions drawn
above by direct inspection of our raw first-principles results, and
also yield a number of additional insights.

First, the strength of the AFD instabilities for small-$t$ compounds
is reflected in the large negative values of $\kappa_{r}$ and
$\kappa_{m}$, which get closer to zero (and eventually become
positive) as the tolerance factor increases. Note that, in principle,
a large energy difference between the AFD phases and the cubic
reference might also originate from small anharmonic couplings
$\bar{\alpha}_{r}$ and $\bar{\alpha}_{m}$ [see Eqs.~(\ref{eq:Etet}),
  (\ref{eq:Erho}), (\ref{eq:Eort}), and (\ref{eq:Eo})]. However, these
parameters do not present any marked or systematic variation with $t$,
and remain in the range between 0.1~eV/\AA$^{4}$ and 0.5~eV/\AA$^{4}$
for all investigated compounds.

Second, we find $\kappa_{r} < \kappa_{m}$ for all the investigated
materials, reflecting the fact that the antiphase rotations
constitute stronger structural instabilities of the cubic phase than
their in-phase counterparts. Describing the anharmonic couplings is
not as straightforward. Roughly, we find that the isotropic coupling
constants $\bar{\alpha}_{r}$ and $\bar{\alpha}_{m}$ are similar for
all the considered compounds, and that we generally have
$\bar{\alpha}_{r} \gtrsim \bar{\alpha}_{m} > 0$. In contrast, we tend
to have $\bar{\gamma}_{r} < \bar{\gamma}_{m} < 0$, which is consistent
with the dominance of the $a^{-}a^{-}a^{-}$ solution over purely
in-phase or other purely antiphase states.

Third, while we obtain $\kappa_{r} < 0$ for all the investigated
compounds, we find two materials (LaAlO$_{3}$ and BaZrO$_{3}$) for
which $\kappa_{m} \gtrsim 0$. In such cases the in-phase tilts are not
instabilities of the cubic phase, and it is thus natural that
structures with only in-phase tilts cannot be stabilized, as mentioned
above. Hence, our usual fitting procedure does not allow us to compute
$\kappa_{m}$ for these compounds; instead, we obtain it by
diagonalizing the Hessian matrix -- of second derivatives of the
energy -- corresponding to the cubic reference structure. Also, as can
be seen in the Tables, for LaAlO$_{3}$ and BaZrO$_{3}$ we do not
compute any anharmonic terms involving in-phase tilts, or the
couplings with strains.

Fourth, our calculated parameters allow us to discuss in detail the
reasons why the ${\cal O}$ phase turns out to be the ground state of
most perovskite oxides. As already mentioned, for all the considered
compounds, antiphase tilts render more stable structures than in-phase
rotations. Further, our fitted PES clearly indicates that the
antiphase and in-phase modes compete with each other, as we get
$\bar{\alpha}_{\rm int}, \bar{\beta}_{\rm int} > 0$ for all studied
materials. Hence, it is now clear that the $a^{-}a^{-}c^{+}$ ground
state, which combines antiphase and in-phase tilts, does {\em not}
emerge because of a cooperation between the two types of AFD
modes. Rather, the ${\cal O}$ phase prevails in spite of the fact that
these two distortions compete and tend to cancel each other.

Let us emphasize this point. Our results clearly show that there is no
such thing as a driving force for the simultaneous occurrence of
antiphase and in-phase tilts in {\sl AB}O$_{3}$ perovskites. Instead,
the reason why they appear together in most compounds is somewhat
mundane. Indeed, all the investigated $Pbnm$ materials share the
feature that $\kappa_{r} \lesssim \kappa_{m} < 0$, i.e., they posses
similarly strong antiphase and in-phase instabilities of the
high-symmetry cubic structure. Thus, in principle such distortions
should occur simultaneously, unless their competition is large enough
for the strongest (${\bf r}$) to suppress the weakest (${\bf m}$). Our
results show that the ${\bf r}$-${\bf m}$ competition is not as
strong, and thus the two tilt types coexist.

\begin{figure}
  \includegraphics[width=0.95\linewidth]{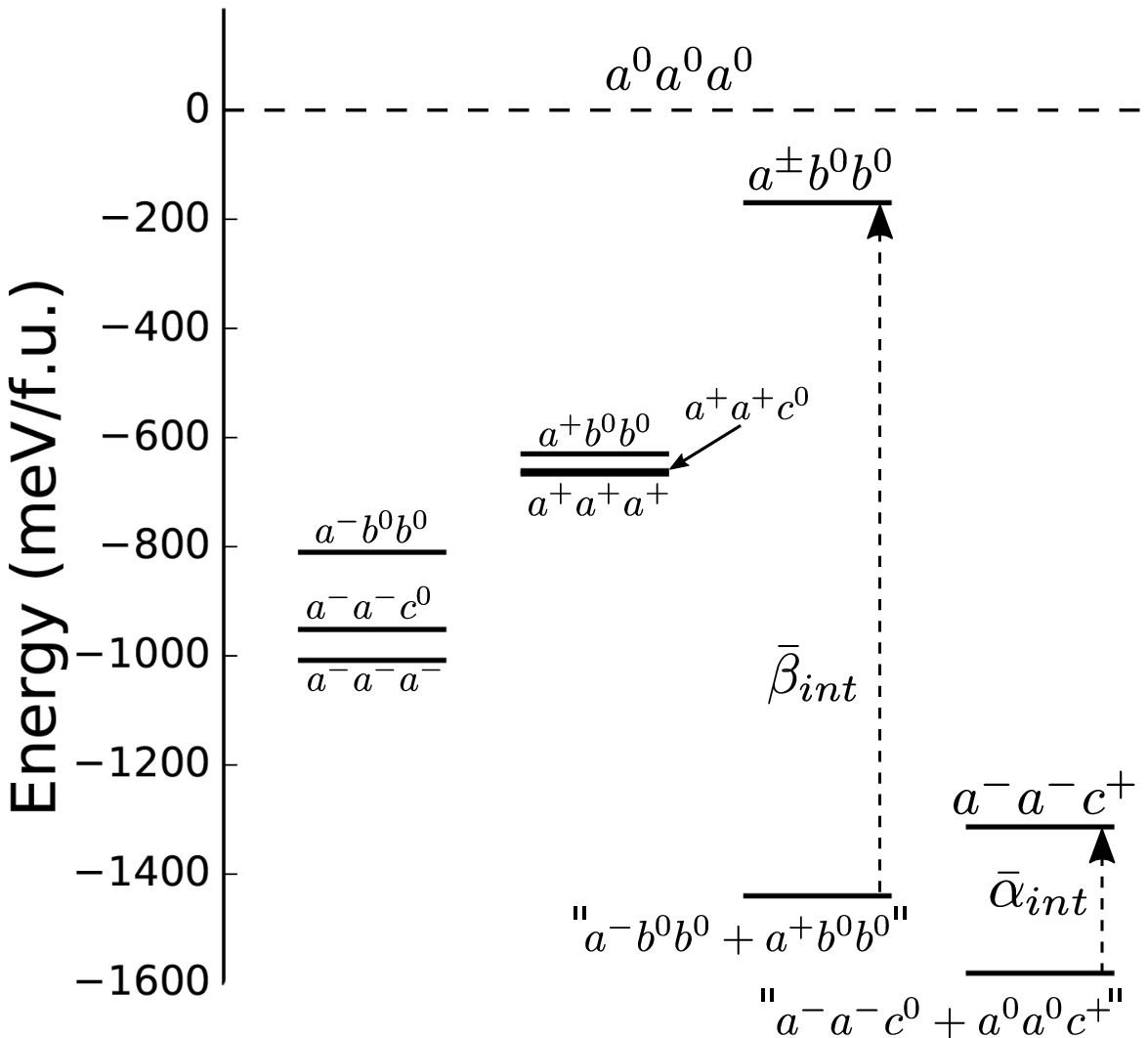}
\caption{Diagram showing the energies of different polymorphs, as well
  as some hypothetical structures, for representative compound
  GdFeO$_{3}$ (see text). The reference cubic phase (denoted by
  $a^{0}a^{0}a^{0}$) is taken as zero of energy. The interactions that
  dominate some energy variations are indicated for emphasis. See text
  for details.}
\label{fig:sketch-energies}
\end{figure}

To gain additional insight, and to understand why the antiphase and
in-phase tilts appear in the specific $a^{-}a^{-}c^{+}$ combination,
let us turn our attention to Fig.~\ref{fig:sketch-energies}. The
diagram shows the relative stability, with respect to the cubic
($a^{0}a^{0}a^{0}$) phase, of different AFD polymorphs for the
representative case of GdFeO$_{3}$. The dominant antiphase-tilted
phase is $a^{-}a^{-}a^{-}$, closely followed by the $a^{-}a^{-}c^{0}$
structure, which lies about 50~meV/f.u. above it. The structures with
only in-phase tilts are about 300~meV/f.u. above the corresponding
antiphase-tilted ones, and the energy gap between the most
($a^{+}a^{+}a^{+}$) and least ($a^{+}b^{0}b^{0}$) stable one is about
35~meV/f.u. Now, for the sake of the argument, let us imagine that the
${\bf r}$ and ${\bf m}$ tilts do not interact. In that case, hybrid
${\bf r}$-${\bf m}$ states like those indicated in
Fig.~\ref{fig:sketch-energies} -- namely,
``$a^{-}b^{0}b^{0}$+$a^{+}b^{0}b^{0}$'' and
``$a^{-}a^{-}c^{0}$+$a^{0}a^{0}c^{+}$''-- could potentially be the
ground state of the material. Indeed, in absence of ${\bf r}$-${\bf
  m}$ interactions, these two structures can be viewed as a simple
combination of antiphase and in-phase distortions, and their energies
with respect to the cubic reference would be $E_{r}^{\rm
  tet}+E_{m}^{\rm tet}-2E_{0}$ and $E_{r}^{\rm ort}+E_{m}^{\rm tet} -
2E_{0}$, respectively. As shown in Fig.~\ref{fig:sketch-energies}, for
GdFeO$_{3}$ this yields energies well below that of the
$a^{-}a^{-}a^{-}$ phase, simply because the energy gain associated to
the condensation of an in-phase tilt ($E_{m}^{\rm tet}-E_{0}$ in these
examples, which is about 630~meV/f.u.) is much greater than the energy
cost of rotating the antiphase-tilt axis (i.e., the anisotropy for
${\bf r}$, as given by $E_{r}^{\rm tet}-E_{r}^{\rm rho} \approx
200$~meV/f.u., is comparatively small). As long as such a condition is
met, having a ground state combining ${\bf r}$ and ${\bf m}$ tilts is
in principle possible.

However, antiphase and in-phase tilts do interact repulsively
($\bar{\alpha}_{\rm int}$, $\bar{\beta}_{\rm int} > 0$), which will
increase the energy of our hypothetical hybrid configurations. The
``$a^{-}b^{0}b^{0}$+$a^{+}b^{0}b^{0}$'' structure will be most
strongly affected, as the occurrence of ${\bf r}$ and ${\bf m}$ tilts
about the same axis is much penalized by the large coupling
$\bar{\beta}_{\rm int}$. In Fig.~\ref{fig:sketch-energies} we show the
energy of such a phase once the ${\bf r}$-${\bf m}$ interactions are
considered; the result, which we denote $a^{\pm}b^{0}b^{0}$, is
obviously not competitive with other polymorphs. In contrast, for
``$a^{-}a^{-}c^{0}$+$a^{0}a^{0}c^{+}$'' the active ${\bf r}$-${\bf m}$
interaction is given by the relatively small $\bar{\alpha}_{\rm int}$
coupling, and the resulting structure ($a^{-}a^{-}c^{+}$) is obviously
competitive with the other low-lying polymorphs. In fact, this is the
$Pbnm$ ground state in the case of GdFeO$_{3}$.

By inspecting the parameters computed in the present investigation, it
is apparent that the above picture applies to all the compounds with a
$Pbnm$ ground state considered in this work. Hence, we think this
picture is likely to be valid for most $Pbnm$ perovskite oxides.

Finally, let us turn our attention to the materials that do not
present a $Pbnm$ ground state. In the case of the large-$t$ compounds,
the situation is quite obvious from the above description. Whenever
$\kappa_{m} > 0$ (LaAlO$_{3}$, BaZrO$_{3}$), there is actually no
driving force for the occurrence of in-phase tilts, and the $Pbnm$
polymorph does not exist. Whenever we have a negative but small
$\kappa_{m}$, we get a $Pbnm$ polymorph that barely differs from a
$a^{-}a^{-}c^{0}$ structure [see Fig~\ref{fig:rawresults}(g)]. In such
cases, the repulsion $\bar{\alpha}_{\rm int}$ is often able to push up
the energy of the ${\cal O}$ phase and yield a purely antiphase-tilted
ground state. We should note that the result may depend on very tiny
energy differences in some limit cases; see e.g. the occurrence of a
$Pbnm$ ground state in NaTaO$_{3}$ ($\kappa_{m} =
-0.4708$~eV/\AA$^{2}$, $\bar{\alpha}_{\rm int} =
0.0693$~eV/\AA$^{4}$), and its absence in NdAlO$_{3}$ ($\kappa_{m} =
-0.3255$~eV/\AA$^{2}$, $\bar{\alpha}_{\rm int} =
0.1203$~eV/\AA$^{4}$). Yet, the general trends are clear.

In the case of the large-$t$ Zn-based compounds, the reasons why we
obtain a lowest-lying $a^{-}a^{-}a^{-}$ state, instead of the fully
developed $a^{-}a^{-}c^{+}$ polymorph, are totally different. Indeed,
by inspecting the parameters in
Table~\ref{tab:strain-renorm-parameters}, we find that these materials
display the following distinct features affecting the
$a^{-}a^{-}a^{-}$ {\sl vs} $a^{-}a^{-}c^{+}$ competition. On one hand,
they present very strong ${\bf r}$-${\bf m}$ repulsive interactions,
featuring record values for $\bar{\alpha}_{\rm int}$ in the case of
ZnTiO$_{3}$ and ZnGeO$_{3}$. On the other hand, they display, by far,
the strongest anisotropies among the investigated compounds, as
quantified by $\bar{\gamma}_{r}$ and $\bar{\gamma}_{m}$. As a result,
antiphase tilts about $\langle 110 \rangle$ and $\langle 100 \rangle$
axes are strongly penalized compared to the $a^{-}a^{-}a^{-}$
state. The combination of these two factors, particularly the latter,
explains why these materials prefer the $R\bar{3}c$
polymorph. Interestingly, a (huge) antiphase rotation with ${\bf r}
\parallel [111]$ constitutes the structural path connecting the
perovskite and LiNbO$_{3}$-type structures. Further, we know that, in
reality, the considered Zn-based compounds crystallize in the
LiNbO$_{3}$-type phase or the (related) ilmenite structure. Hence, our
present results in the small-$t$ limit -- featuring $\kappa_{r} \ll 0$
and $\bar{\gamma}_{r} \ll 0$ -- reflect the well-known tendency to
abandon the perovskite lattice and move towards a LiNbO$_{3}$-like
structure.\cite{benedek13} As a by-product of sorts, the ${\cal O}$
phase losses its predominance in this limit.

Note that the above observations can be confirmed by considering the
formulae in Section~\ref{sec:formalism} and the actual parameters
obtained for specific materials. For example, it is straightforward to
check why the energy of the ${\cal O}$ phase [$E^{\cal O}$,
  Eq.~(\ref{eq:Eo})] will be generally lower than that of competing
polymorphs: it benefits from the contributions from both antiphase and
in-phase distortions, while $\bar{\alpha}_{\rm int}$ is relatively
small. It is also easy to understand why the $Pbnm$ solution is a
minimum of the energy [Eq.~(\ref{eq:hessian-elements-ophase})], as
this is essentially guaranteed by the positive interaction terms
$\bar{\alpha}_{\rm int}$ and $\bar{\beta}_{\rm int}$. Further, it can
be readily seen that, whenever $\kappa_{m} > 0$, the possible
existence of the ${\cal O}$ phase as a singular point is unclear, as
we would typically have $(m^{\cal O})^{2} < 0$ in
Eq.~(\ref{eq:ophase-m}) if all the other parameters have values as
those computed here. It is true that, from Eq.~(\ref{eq:ophase-m}),
one might imagine alternative ways to stabilize the ${\cal O}$ phase
even if $\kappa_{m}$ is positive; for example, we might have a strong
cooperative interaction $\bar{\alpha}_{\rm int} < 0$, while keeping
$\kappa_{r} < 0$ as the main driving force for the structural
instability. Nevertheless, according to our DFT results, all the
investigated compounds are far from such alternative scenarios, which
thus seem to be highly unlikely.

\subsection{Strain effects}

Let us now turn our attention to the elastic energy ($E_{s}$) and the
coupling of strains with the tilt modes ($E_{sp}$). The corresponding
parameters are given in Table~\ref{tab:bare-parameters}, as obtained
from the fit of all the bare coupling constants in
Eq.~(\ref{eq:energy}).

As regards $E_{s}$, the behavior of the investigated materials is
standard, the cubic phase being stable against strains. It is
interesting to note that there is no clear dependence of the elastic
constants on the tolerance factor, suggesting that chemical
considerations -- as opposed to steric -- should be most relevant in
this case.

As regards the coupling between strains and AFD modes, we find that
the constants $B_{r1xx}$, $B_{r1yy}$, $B_{m1xx}$, and $B_{m1yy}$ are
positive for some compounds and negative for others. Thus, for
example, LaAlO$_{3}$ presents negative values of $B_{r1xx}$ and
$B_{r1yy}$, implying that negative strains $\eta_{1}$, $\eta_{2}$,
$\eta_{3} < 0$ -- i.e., a smaller cell volume -- will tend to weaken
the ${\bf r}$ instabilities; this is compatible with the known
behavior of LaAlO$_{3}$, as it is experimentally and computationally
observed that an hydrostatic compression results in a transition from
the usual tilted phase of the compound ($R\bar{3}c$,
$a^{-}a^{-}a^{-}$) to a non-tilted structure (cubic
$Pm\bar{3}m$).\cite{bouvier02} In contrast, positive values of these
strain-tilt couplings imply the opposite effect, that is, an
enhancement of the rotational instabilities upon compression; this is
the most common behavior, as discussed at length by some of us in
Ref.~\onlinecite{xiang17}. In addition, we find that the coupling
constant between shear strains and antiphase rotations ($B_{r4yz}$)
varies sign depending on the compound. Finally, it seems all but
impossible to identify clear trends of the strain-phonon coupling
parameters as a function of tolerance factor, which suggests that
other (chemical) factors must play a role in determining their
value. This issue, which is the focus of ongoing studies by some of
us,\cite{gu17} falls beyond the scope of this work and will not be
pursued here.

Rather, our present interest is to understand how strain affects the
relative stability of the tilt phases. To gain insight into this
question, we show in Fig.~\ref{fig:parameters} the most important bare
parameters [e.g., $\alpha_{r}$, $\gamma_{r}$, etc., obtained by
  fitting Eq.~(\ref{eq:energy}) to our DFT results] together with
their strain-renormalized counterparts [e.g., $\bar{\alpha}_{r}$,
  $\bar{\gamma}_{r}$, etc., obtained by fitting
  Eq.~(\ref{eq:renorm-energy})]. Note that a difference between bare
and strain-renormalized couplings is indicative of a strain
relaxation. Our main findings are as follows.

First, for the harmonic parameters $\kappa_{r}$ and $\kappa_{m}$, we
obtain essentially the same values from the two fitting procedures,
for all investigated compounds. This is the expected result because,
provided our fourth-order series is an accurate representation of the
relevant PES, we should not have any strain renormalization of the
harmonic constants (see Section~\ref{sec:strain-free}). Second, the
strain renormalization is also negligible for the interaction
couplings, so that we have $\bar{\alpha}_{\rm int} \approx \alpha_{\rm
  int}$ and $\bar{\beta}_{\rm int} \approx \beta_{\rm int}$. This
result is not obvious {\sl a priori}, and indicates that, for the
investigated compounds, strain does not play any significant role in
the competition between antiphase and in-phase rotations. Third, there
is a sizeable renormalization of the $\alpha_{r}$ and $\alpha_{m}$
parameters for some of the compounds studied (e.g., NaTaO$_{3}$),
although the effect has no qualitative significance. Note that we
always have $0 < \bar{\alpha}_{r} \lesssim \alpha_{r}$ and $0 <
\bar{\alpha}_{m} \lesssim \alpha_{m}$, i.e., the strain results in
larger tilt distortions by weakening the anharmonic (repulsive)
interaction. This is easy to understand: For given values of ${\bf r}$
and ${\bf m}$, the energy for fixed (zero) strains will be higher than
the one obtained if we allow the strains to relax in response to the
tilts. The former case is captured by the bare couplings, and the
latter by the strain-renormalized ones; the mentioned energy reduction
corresponds to having $\bar{\alpha}_{r}$ and $\bar{\alpha}_{m}$
strictly smaller than $\alpha_{r}$ and $\alpha_{m}$,
respectively. Finally, the anisotropy terms $\gamma_{r}$ and
$\gamma_{m}$ also exhibit a significant strain renormalization for
some compounds, although the effect is generally small. In this case,
we have no definite expectations on the behavior of the renormalized
parameters and, indeed, our findings do not show any obvious
systematics. It is worth noting that, in cases in which $\gamma_{r}$
or $\gamma_{m}$ is close to zero, the strain relaxation may cause the
coupling to change sign, and thus reverse the relative stability of
the tetragonal (e.g., $a^{-}b^{0}b^{0}$) and rhombohedral (e.g.,
$a^{-}a^{-}a^{-}$) structures (see
Section~\ref{sec:singular}). According to our results, SrGeO$_{3}$
presents this behavior ($\gamma_{r} = -0.193$~eV/\AA$^{4}$,
$\bar{\gamma}_{r} = 0.004$~eV/\AA$^{4}$), and NdAlO$_{3}$ and
PrAlO$_{3}$ are borderline cases. This extreme sensitivity to strain
is best characterized theoretically in ferroelectric
PbTiO$_{3}$,\cite{kingsmith94,wojdel13} and our results here provide
an AFD analogue of such an effect.

Hence, while strains do have some impact on our investigated PES, the
effects are of little importance to the central question here, i.e.,
the preeminence of the $Pbnm$ structure among perovskites. Indeed,
strain effects -- which are negligible for the interacting constants
$\alpha_{\rm int}$ and $\beta_{\rm int}$ -- are largely irrelevant in
that respect. Let us note that we corroborated this conclusion by
repeating the computational investigation of our thirty-five
compounds, considering all the AFD polymorphs mentioned above, under
the constraint of zero strains. (We thus impose that the lattice
vectors be fixed at the values obtained from the symmetry-constrained
relaxation of the cubic structure.) By fitting
Eq.~(\ref{eq:renorm-energy}) to the DFT data thus computed, we obtain
parameters that are qualitatively identical, and quantitatively very
similar, to our strain-renormalized results in
Table~\ref{tab:strain-renorm-parameters}. Hence, strains will not be
further considered here.

\subsection{{\sl A}-site antipolar distortions}

Antipolar displacements of the {\sl A} cations, as those shown in
Fig.~\ref{fig:sketch-dists}, have been found to play an important role
in stabilizing the $a^{-}a^{-}c^{+}$ structure over competing
polymorphs in some compounds.\cite{benedek13,dieguez11,miao14} In this
Section we discuss how such modes can be treated, and their effect
quantified and analyzed, within our present scheme.

\begin{figure*}
\includegraphics[width=1.00\linewidth]{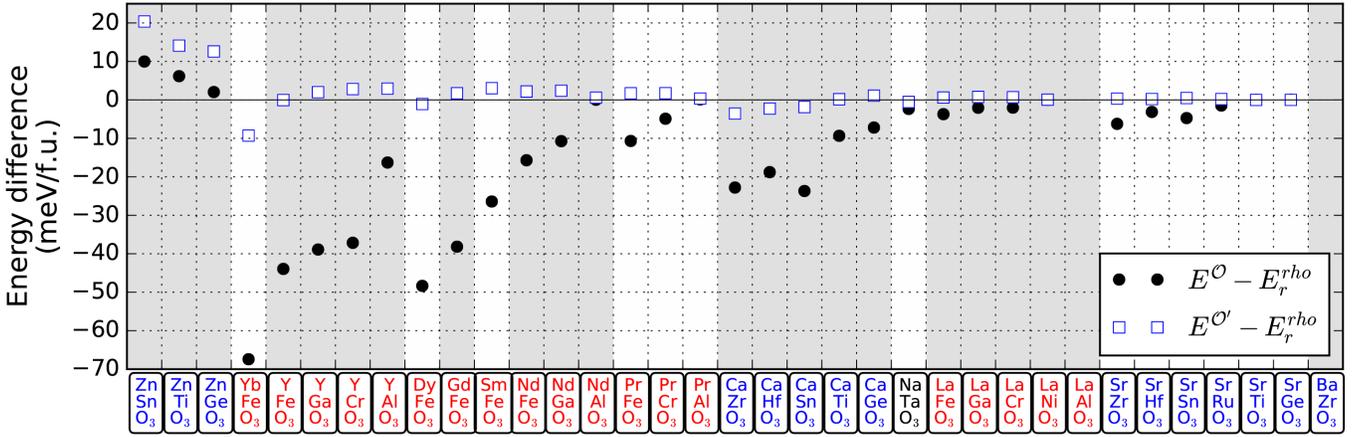}
\caption{Energy difference (meV/f.u.) between the $Pbnm$ and
  $R\bar{3}c$ phases, as obtained in usual (filled circles, $E^{\cal
    O} - E^{\rm rho}_{r}$) and frozen-{\sl A} (open squares, $E^{{\cal
      O}'} - E^{\rm rho}_{r}$) conditions. A negative energy
  difference implies that the $Pbnm$ phase is more stable. The color
  code for the chemical formulas is as in Fig.~\ref{fig:rawresults}.}
\label{fig:frozenA}
\end{figure*}

Let us first test the importance of the antipolar distortions by
performing the following computational experiment: For all the
materials considered here, we repeat the relaxation of the $Pbnm$
structure under the constraint that the {\sl A}-cations be frozen in
their high-symmetry positions. In other words, we impose null
antipolar distortions and thus preclude the possibility that the {\sl
  A}-cations may move off-center to optimize the energy of the ${\cal
  O}$ phase. In the following we will refer to such a constraint as
``frozen-{\sl A}'', and the quantities computed in frozen-{\sl A}
conditions will be primed. Figure~\ref{fig:frozenA} summarizes our
findings, showing how the energy difference between the
$a^{-}a^{-}c^{+}$ and $a^{-}a^{-}a^{-}$ structures varies depending on
whether the antipolar modes are allowed or not. The results are
crystal clear: We observe that, for most of the considered compounds,
the $Pbnm$ and $R\bar{3}c$ phases become nearly degenerate in
frozen-{\sl A} conditions. Further, we typically have $E_{r}^{\rm rho}
\lesssim E^{{\cal O}'}$, so that the $R\bar{3}c$ structure becomes the
lowest-energy state. Hence, previous observations in the literature
get confirmed: the antipolar {\sl A}-cation distortions are essential
for the preeminence of the $Pbnm$ ground state among perovskite
oxides. In their absence, most compounds would present an
$a^{-}a^{-}a^{-}$ ground state.

These antipolar distortions can be thought of as secondary modes that
follow the primary ${\bf r}$ and ${\bf m}$ order parameters in the
same way that strain does. Hence, the antipolar modes are naturally
present whenever we relax the $Pbnm$ phase for any of the considered
compounds; they couple to the octahedral tilts and strains, and thus
contribute to the resulting structure and energy of the ${\cal O}$
phase. Consequently, the effect of these modes is implicitly captured
when we fit the parameters describing the relevant PES to DFT
data. Note that this interpretation of the antipolar distortions as
secondary modes is strictly correct only for compounds that do not
present antipolar instabilities of the cubic phase, as is the case of
the majority of materials here considered (see discussion in
Section~\ref{sec:methods} and below). It is also important to realize
that, from all the AFD polymorphs discussed above, such antipolar
distortions appear only in the $a^{-}a^{-}c^{+}$ structure; in fact,
it can be checked that none of the simpler phases considered here, for
which we have either ${\bf m} = {\bf 0}$ or ${\bf r} = {\bf 0}$,
presents any secondary distortions besides strain.\cite{bellaiche13}

Since the antipolar distortions are treated implicitly in our PES
description, we can view our calculated parameters in
Tables~\ref{tab:bare-parameters} and
\ref{tab:strain-renorm-parameters} as being renormalized by these
modes. Nevertheless, we can go further and explicitly study such a
renormalization by recalling the relevant couplings between antipolar
and AFD modes, which have been discussed elsewhere.\cite{bellaiche13}
For simplicity, in the following we consider the particular
$a^{-}a^{-}c^{+}$ variant of the ${\cal O}$ phase, noting that the
results for other, equivalent structures -- e.g., the one described by
$a^{-}b^{+}a^{-}$ -- can be obtained directly by suitable symmetry
transformations.

There are two antipolar modes associated to off-centering
displacements on the {\sl A}-cations. In the following we discuss at
length the first and most relevant of them, which yields the largest
structural distortions and associated energy reductions. We will
briefly discuss the second one at the end of the Section.

The first antipolar mode features movements of the {\sl A} cations
along the [110] pseudo-cubic direction, spatially modulated according
to the ${\bf q}_{Z} = \pi/a(0,0,1)$ wave vector [see
  Fig.~\ref{fig:sketch-dists}(c)]. This mode involves an homogeneous
pattern of [110]-oriented dipoles in a given (001) plane, and the
reversal of such dipoles as we move by one elemental cell along the
[001] direction. Let $Z_{xy}$ be the amplitude of this
distortion. Following Ref.~\onlinecite{bellaiche13}, one can prove
that its lowest-order coupling with the AFD modes has the form
\begin{equation}
E_{{\rm int}Z} = \beta_{{\rm int}Z} Z_{xy}r_{xy}m_{z} \; ,
\end{equation}
where $\beta_{{\rm int}Z}$ is a material-dependent constant and we
assume that the $a^{-}a^{-}c^{+}$ state is characterized by
\begin{equation}
\begin{split}
  {\bf r} =& \; r_{xy} (1,1,0) \\
  {\bf m} =& \; m_{z}  (0,0,1) \; .
\end{split}
\end{equation}
Let the energy associated to this antipolar mode be given by
\begin{equation}
E_{Z} = \kappa_{Z} Z_{xy}^{2}
\end{equation}
with $\kappa_{Z} > 0$, as it corresponds to a regular distortion that
is not an instability of the cubic phase. [If $\kappa_{Z}$ were
negative, we would need to introduce ${\cal O}(Z_{xy}^{4})$ terms in
$E_{Z}$.]  We can add $E_{Z}$ and $E_{{\rm int}Z}$ to the energy in
Eq.~(\ref{eq:energy}) and, in analogy to our treatment for the strain
in Section~\ref{sec:strain-free}, impose the equilibrium condition
\begin{equation}
\left.\frac{\partial E}{\partial Z_{xy}}\right|_{\rm eq} = 0 \, ,
\end{equation}
which yields the distortion
\begin{equation}
Z^{\rm eq}_{xy} = - \frac{\beta_{{\rm int}Z}}{2\kappa_{Z}} \; r_{xy}m_{z}
\; .
\end{equation}
If we substitute this result into the above expressions for $E_{Z}$
and $E_{{\rm int}Z}$, we obtain
\begin{equation}
E_{Z} + E_{{\rm int}Z} = - \frac{\beta^{2}_{{\rm
      int}Z}}{4\kappa_{Z}} \; r_{xy}^{2}m_{z}^{2} \; ,
\end{equation}
where there is no explicit dependence on the antipolar mode
amplitude. Now, by recalling the form of the energy for an
$a^{-}a^{-}c^{+}$ state [Eq.~(\ref{eq:energy-ophase-raw})], we can see
that the coupling term stemming from $E_{Z}+E_{{\rm int}Z}$
contributes exclusively to the anharmonic interaction constant
$\bar{\alpha}_{\rm int}$. [If we work with the full expression for the
  energy (Eq.~\ref{eq:energy}), we trivially find that the
  renormalized anharmonic coupling is $\alpha_{\rm int}$. Further, if
  we write the full symmetry invariant for the trilinear ${\bf
    Z}$-${\bf r}$-${\bf m}$ coupling, we obtain a renormalization term
  proportional to $r^{2}m^{2}$, which contributes to $\alpha_{\rm
    int}$ in Eq.~(\ref{eq:int-energy}).]

It is important to note that this contribution to $\bar{\alpha}_{\rm
  int}$ is negative. In other words, the $Z_{xy}$ relaxation favors an
attractive, cooperative anharmonic interaction between antiphase and
in-phase tilts. As a consequence, it tends to stabilize structures
that combine both types of tilts about certain specific axes (e.g.,
${\bf r} \parallel [110]$ and ${\bf m} \parallel [001]$ in our case),
and will result in larger tilt amplitudes and a lower energy $E^{\cal
  O}$.

We can test this theoretical prediction numerically. As mentioned
above, we have DFT results for relaxed ${\cal O}$ phases in absence of
antipolar distortions (frozen-{\sl A} conditions). Hence, we can use
those data, together with our DFT results for the simpler
only-antiphase and only-in-phase AFD states, to compute the coupling
constants that describe the corresponding PES. The main outcome of
this exercise is shown in Fig.~\ref{fig:parameters}, where the effect
of the antipolar renormalization on the $\bar{\alpha}_{\rm int}$
parameters is clearly visible. (We get no significant difference for
the other coupling constants, in agreement with the theoretical
expectations.) Indeed, for all compounds we find $\bar{\alpha}'_{\rm
  int} > \bar{\alpha}_{\rm int} > 0$, where a larger
$\bar{\alpha}'_{\rm int}$ implies a greater ${\bf r}$-${\bf m}$
competition. As shown in Fig.~\ref{fig:frozenA}, such a competition
can become strong enough as to yield an $a^{-}a^{-}a^{-}$ ground
state.

In view of these findings, we can conclude that the preeminence of the
$Pbnm$ ground state over the $R\bar{3}c$ polymorph stems from a
balance between the tendency of the material to condense both
antiphase and in-phase tilts ($\kappa_{r} < \kappa_{m} < 0$) and the
mutually-exclusive interaction between them ($\bar{\alpha}'_{\rm int}
> 0$).  This balance is a delicate one. Indeed, as shown in
Fig.~\ref{fig:frozenA}, it typically involves small energy differences
$|E^{{\cal O}'}-E_{r}^{\rm rho}| \approx 10$~meV/f.u., the
$a^{-}a^{-}a^{-}$ phase being dominant in frozen-{\sl A}
conditions. Then, the extra energy reduction provided by the
relaxation of antipolar modes ($E^{\cal O}-E^{{\cal O}'}$) is usually
enough to tip the balance and stabilize the ${\cal O}$ ground state.

Finally, let us comment on the second antipolar mode occurring in the
$Pbnm$ phase [Fig.~\ref{fig:sketch-dists}(d)], which involves
displacements of the {\sl A}-cations along the $[1\bar{1}0]$
pseudo-cubic direction, modulated according to the ${\bf q}_{R}$ wave
vector. Following Ref.~\onlinecite{bellaiche13}, we know that the
leading coupling responsible for the activation of this secondary mode
has the form
\begin{equation}
E_{{\rm int}R} = \beta_{{\rm int}R} R_{x\bar{y}}r_{xy}m_{z}^{2} \; ,
\end{equation}
where $R_{x\bar{y}}$ is the amplitude of the ${\bf q}_{R}$-modulated
antipolar distortion and $\beta_{{\rm int}R}$ is a material-dependent
coupling constant. Assuming that the energy of this mode is given by
\begin{equation}
E_{R} = \kappa_{R} R_{x\bar{y}}^{2} 
\end{equation}
with $\kappa_{R} > 0$, the tilt-dependent equilibrium value of
$R_{x\bar{y}}$ is
\begin{equation}
R^{\rm eq}_{x\bar{y}} = - \frac{\beta_{{\rm int}R}}{2\kappa_{R}} \; r_{xy}m_{z}^{2}
\; ,
\end{equation}
and its contribution to the energy is
\begin{equation}
E_{R} + E_{{\rm int}R} = - \frac{\beta^{2}_{{\rm
      int}R}}{4\kappa_{R}} \; r_{xy}^{2}m_{z}^{4} \; .
\end{equation}
This result is similar to the one obtained above for the $Z_{xy}$
distortion. In fact, the qualitative effect of this second antipolar
renormalization -- i.e., to favor the simultaneous occurrence of
antiphase and in-phase tilts -- is exactly the same. There is one
important difference, though: Relaxing the $R_{x\bar{y}}$ mode affects a
sixth-order interaction between the tilts, a coupling that is not
included in our fourth-order model of the relevant PES. Since our
numerical results regarding the $Pbnm$ {\sl vs} $R\bar{3}c$
competition seem perfectly consistent with a fourth-order Taylor
series, we can conclude that the effect of this second antipolar
renormalization is probably small. Hence, we do not pursue this issue
further in this work.

\subsection{Additional remarks}
\label{sec:remarks}

Let us conclude with some additional comments on our results.

\begin{figure}
\includegraphics[width=0.90\linewidth]{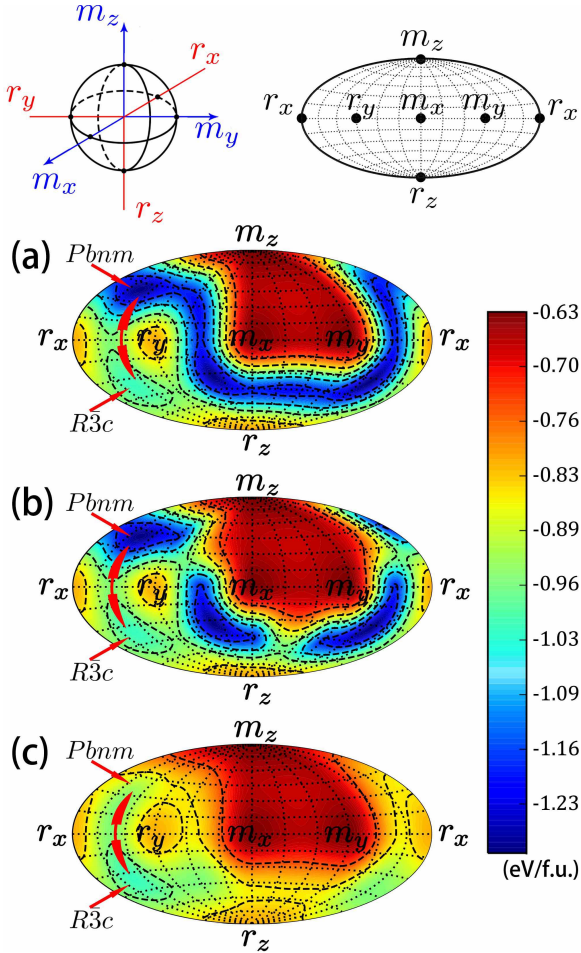}
\caption{Stereographic projection of the strain-renormalized PES of
  GdFeO$_{3}$, as given by $\bar{E}({\bf r},{\bf m})$. Panels~(a) and
  (b) show the actual PES, as obtained from the usual 4th-order and
  corrected 6th-order models, respectively. Panel~(c) displays the
  result obtained in frozen-{\sl A} conditions, and is derived from a
  model that includes a small 6th-order correction. Note that the
  directions corresponding to pure tilts are marked by their
  respective $r_{\alpha}$ or $m_{\alpha}$ symbols. We also mark the
  $Pbnm$ and $R\bar{3}c$ states. See text for details.}
\label{fig:stereo}
\end{figure}

\subsubsection{Energy landscape, sixth-order corrections}
\label{sec:sixth-order}

The above discussion focuses on the relative stability of the
lowest-energy AFD polymorphs, i.e., the $a^{-}a^{-}a^{-}$ and
$a^{-}a^{-}c^{+}$ phases. Nevertheless, from the PES given by our
fitted $\bar{E}({\bf r},{\bf m})$, we have access to the full
six-dimensional energy landscape and can thus explore its features. To
do this, we find it convenient to implement the special stereographic
projection shown in Fig.~\ref{fig:stereo}. We work with a
three-dimensional Cartesian system, with coordinates ${\boldsymbol
  \psi} = (\psi_{x}, \psi_{y}, \psi_{z})$, in which a positive value
of $m_{\alpha}$ ($r_{\alpha}$) correspond to a positive (negative)
value of $\psi_{\alpha}$. We can further define the two-dimensional
surface obtained by minimizing the energy along the radial coordinate
$\psi = |{\boldsymbol \psi}|$, and make a stereographic projection of
the result. We can thus inspect the PES regions in which the energy is
lowest. Figure~\ref{fig:stereo}(a) shows the energy map thus obtained
for representative compound GdFeO$_{3}$.

Before commenting on the features of this landscape, let us note the
low-energy (dark blue) path displayed by Fig.~\ref{fig:stereo}(a),
which connects the following string of structures: $r_{x}r_{y}m_{z}
\rightarrow r_{y}m_{z} \rightarrow m_{x}r_{y}m_{z} \rightarrow
m_{x}r_{y} \rightarrow m_{x}r_{y}r_{z}$, where we start from the
$a^{-}a^{-}c^{+}$ ($Pbnm$) phase indicated with an arrow in the figure
and, as we move to the right and down, end up in an equivalent
$a^{+}b^{-}b^{-}$ structure. According to our fitted 4th-order PES,
all the structures along this path are rather low in energy; in
particular, the $a^{+}b^{-}a^{+}$ phase is predicted to the second
most stable polymorph of GdFeO$_{3}$, only behind the $Pbnm$ ground
state, and lying lower than the $R\bar{3}c$ phase. This is a
surprising result, as the $a^{+}b^{-}a^{+}$ tilt pattern is quite rare
in nature; hence, we run first-principles simulations to verify
it. Interestingly, the DFT simulations reveal that our 4th-order model
-- fitted to account for the $Pbnm$--$R\bar{3}c$ competition, as
explained above -- exaggerates the stability of the $a^{+}b^{-}a^{+}$
polymorph by about 240~meV/f.u.; in fact, we find that, at the DFT
level, the $a^{+}b^{-}a^{+}$ phase lies above the $a^{-}a^{-}a^{-}$
structure by about 40~meV/f.u.

This result indicates that our 4th-order model is not sufficient to
account for the details of the ${\bf r}$-${\bf m}$ interactions in a
quantitatively accurate way. In hindsight, this finding is not
surprising. For GdFeO$_{3}$, and for most of the compounds considered
here, the tilt amplitudes are very large, and it is natural for
couplings above 4th-order to play a role. Specifically, our DFT
result for the $a^{+}b^{-}a^{+}$-type structures can be easily
reproduced by extending the model with an additional 6th-order
coupling of the form
\begin{equation}
\Delta E_{\rm int} = \bar{\gamma}_{\rm int} (r^{2}_{x}m^{2}_{y}m^{2}_{z} +
r^{2}_{y}m^{2}_{z}m^{2}_{x} + r^{2}_{z}m^{2}_{x}m^{2}_{y} ) \; ,
\end{equation}
which has the peculiarity of having no effect at all on the energy and
stability of all the polymorphs discussed above. (For the
$\bar{\gamma}_{\rm int}$ coupling to be active, at least two in-phase
tilt components must be different from zero.) We find that for
$\bar{\gamma}_{\rm int} = 0.2706$~eV/\AA$^{6}$ we recover the DFT
result for the energy of GdFeO$_{3}$'s $a^{+}b^{-}a^{+}$
structure. The corrected energy $\bar{E}$ yields the landscape shown
Fig.~\ref{fig:stereo}(b). The new map is overall quite similar to that
of Fig.~\ref{fig:stereo}(a), except that the $a^{+}b^{-}a^{+}$-like
phases are relatively high-energy saddle points now.

It is apparent from this map that the $Pbnm$ and $R\bar{3}c$ states
both constitute energy minima, and that such minima are connected by a
low-energy $a^{-}a^{-}c^{0}$ saddle point. Simple tetragonal
structures, like those corresponding to the points marked $r_{z}$ and
$m_{z}$, are relatively high-energy saddles that appear as local
maxima in our projection (for such states the energy is convex only
along the radial direction $|{\boldsymbol \psi}|$); in contrast, the
orthorhombic structures $a^{-}a^{-}c^{0}$ and $a^{+}a^{+}c^{0}$ are
lower-energy saddles, reflecting that we have
$\bar{\gamma}_{r},\bar{\gamma}_{m} < 0$ for this material. A peculiar
case is that of the $a^{+}b^{-}c^{0}$ state, e.g., the mid point
between $m_{x}$ and $r_{y}$ in Fig.~\ref{fig:stereo}(b): Note that an
$r_{z}$-distortion reduces the energy of such a structure, as it takes
it towards the $Pbnm$ ground state; in contrast, an $m_{z}$-distortion
increases the energy, as in this case the condensation of a new
in-phase rotation ($\kappa_{m}<0$, $\bar{\gamma}_{m} < 0$) cannot
compensate for the increase in ${\bf r}$-${\bf m}$ repulsion
($\bar{\alpha}_{\rm int}, \bar{\gamma}_{\rm int} > 0$).

In Fig.~\ref{fig:stereo}(c) we show the stereographic projection of
the PES of GdFeO$_{3}$ obtained under the frozen-{\sl A}
constraint. (This PES includes a small 6th-order correction term with
$\bar{\gamma}'_{\rm int} = 0.0192$~eV/\AA$^{6}$.) As compared with the
actual PES [Fig.~\ref{fig:stereo}(b)], the most significant changes
occur in the vicinity of the $Pbnm$ state. Indeed, the increased
$\bar{\alpha}_{\rm int}'$ competition results in the near
disappearance of the $a^{-}a^{-}c^{+}$ minimum and the consolidation
of the $a^{-}a^{-}a^{-}$ ground state. Note also that the energy of
other phases involving {\sl A}-cation relaxations -- e.g., the
$a^{+}b^{-}a^{+}$ and $a^{+}b^{-}c^{0}$ structures\cite{bellaiche13}
-- increases significantly in frozen-{\sl A} conditions, while, in
contrast, the purely antiphase or in-phase states (e.g.,
$a^{-}a^{-}a^{-}$, $a^{+}b^{0}b^{0})$ remain unaltered.

Let us conclude this part by noting that the issue discussed above for
GdFeO$_{3}$ -- i.e., the exaggerated stability of the
$a^{+}b^{-}a^{+}$ structure, as predicted by the default 4th-order
model -- is general among the compounds studied in this work. Indeed,
we used our DFT methods to relax the $a^{+}b^{-}a^{+}$ phase of all
investigated materials, and found that the error in the energy
predicted by the 4th-order model tends to grow as the tolerance factor
decreases. (It can be as large as 400~meV/f.u. for YbFeO$_{3}$.) This
is a reasonable result: smaller-$t$ compounds display larger tilts
and, as a consequence, higher-order energy terms should become more
relevant for an accurate PES description. Following the recipe
described above, we can compute $\bar{\gamma}_{\rm int}$ for all the
investigated compounds that present a stable $a^{+}b^{-}a^{+}$ phase;
the results are given in Table~\ref{tab:strain-renorm-parameters}.

These results show that our fourth-order models -- which
  are sufficient to reproduce the low-lying PES accurately, and whose
  simplicity allows satisfying physical interpretations -- perform
  poorly when it comes to predict the energies of less-favorable
  states. While this seems an acceptable compromise in the present
  study, such an inaccuracy might become a problem if, for example,
  these same models were used to predict the behavior of the materials
  under the action of fields (epitaxial stress \cite{pertsev98},
  electric \cite{stengel15}) that can be expected to stabilize unusual
  phases. Indeed, our results suggest that, generally speaking, one
  should validate low-order potentials before using them for
  quantitative investigations of materials subject to significant
  perturbations.

\subsubsection{More on the phase diagram}

To conclude, let us comment on the scope of the present investigation
as regards a full discussion of the phase diagram of these
perovskites. In this work we compare the relative stability of
different structural phases by inspection of their energies, as
directly obtained from DFT simulations. We focus on discussing the
character of the ground state structure, and are thus confined to the
limit of very low temperatures (strictly speaking, to 0~K). Then, it
is worth noting that, for the prediction of the ground state to be
more accurate, one should add the zero-point contribution to our
computed {\em static} energies. Yet, here we leave zero-point energies
out of the discussion, essentially for two reasons: First, they are
not defined for many of the considered structures, which are saddle
points of the PES and thus have imaginary-frequency phonons associated
to them. Second, they depend on the soft modes of the material as much
as (actually, less than) they depend on the harder ones; hence, the
discussion of zero-point energies has little to do with the PES of the
tilt modes, which is our main focus in this work.

To these main reasons, let us add the expectation that, because all
the AFD phases of a given compound share the same kind of lattice
topology and chemical bonding, zero-point energies should not be
strongly polymorph dependent. It is obvious, though, that we have
materials in which the obtained energy gap between different phases is
tiny, and in such cases zero-point effects might in principle tip the
balance. Nevertheless, the general agreement between our results for
the lowest-energy structure and experimental observations (we are not
aware of any obvious conflict) suggests that zero-point energies do
not play any important role in determining the ground state of these
compounds. Let us note that this seems to be the case of most
first-principles works with perovskite oxides, as zero-point
corrections are seldom considered and, yet, good agreement with
experiment is common.\cite{rabe-book2007}

Finally, let us briefly comment on how one could investigate the
effect of temperature on the competition among different tilt
polymorphs. We have two distinct situations. On one hand, whenever we
have compounds with well-developed O$_{6}$ rotations, for which the
$Pbnm$ and $R\bar{3}c$ structures are local energy minima, it should
be possible to account for the effect of temperature by using the
well-known quasi-harmonic approximation (see, e.g.,
Ref.~\onlinecite{cazorla13}). This would require accurate and heavy
calculations of the phonon spectrum, as a function of volume, from
which the temperature-dependent free energy of the different
polymorphs can be approximated. Such an approach that has been barely
applied in studies of perovskite oxides. Yet, it is interesting to
note that a detailed investigation of BiFeO$_{3}$
(Ref.~\onlinecite{cazorla13}) showed that the $a^{-}a^{-}c^{+}$
structure is softer than the $a^{-}a^{-}a^{-}$ polymorph (which is
polar, with space group $R3c$, for this compound), and it becomes
favored upon heating. Interestingly, if the occurrence of a relatively
soft $Pbnm$ phase were general in the perovskite family, thermal
effects would provide us with yet another reason for the prevalence
of the ${\cal O}$ state. Nevertheless, this point should be explicitly
verified on a case by case basis, as we do not see any general reasons
for the $a^{-}a^{-}c^{+}$ structure to be softer (or harder) than the
$a^{-}a^{-}a^{-}$ one.

On the other hand, for compounds with weak tilt instabilities, the
phase diagram will probably be determined by mechanisms that are
typical of displacive soft-mode transitions. Hence a theoretical
discussion will require a treatment of unstable phonon bands that will
be, presumably, strongly temperature dependent. To study such cases we
would need to resort to effective-potential schemes like e.g. those
introduced in Refs.~\onlinecite{zhong94a,zhong95a,wojdel13}. The few
existing studies applying such methods to AFD compounds suggest that
subtle interactions control the phase
diagram,\cite{wojdel13,zhong95b,akbarzadeh05} which dissuades us from
formulating any general expectations.

\section{Conclusions}
\label{sec:conclusions}

In summary, this article reports on a thorough theoretical
investigation of {\sl AB}O$_{3}$ perovskite oxides whose structure is
characterized by concerted tilts of the O$_{6}$ octahedra that
constitute the backbone of the lattice. Our results
provide a clear picture of why one particular tilt polymorph (the
$a^{-}a^{-}c^{+}$ pattern, corresponding to the orthorhombic $Pbnm$
space group) prevails over all other in most perovskite materials;
indeed, we are able to identify the physical requirements for such a
structure to occur -- i.e., antiphase and in-phase tilts are both
native instabilities of the cubic perovskite prototype, relatively
small anisotropy energy of the antiphase tilts, relatively weak
competition between antiphase and in-phase tilts --, which happen to
occur very frequently. Our results also prove the critical role played
by secondary distortions -- antipolar modes involving the {\sl A}
cations -- to weaken the antiphase/in-phase competition and yield the
$Pbnm$ ground state. Additionally, we find that the $Pbnm$ polymorph
losses its preeminence in two opposite limits -- essentially, for
small and large {\sl A} cations -- for completely different reasons,
which we discuss in some detail. Hence, beyond corroborating some
scattered observations in the literature, this work brings
unprecedented insight into (and quantification of) the competition
between different tilt phases in perovskite oxides, and we hope will
be useful to better understand existing compounds and eventually
design new ones.

This work was mainly funded by the Luxembourg National Research Fund
through Grants FNR/P12/4853155/Kreisel COFERMAT (H.J.Z. and J.\'I.),
FNR/C15/MS/10458889 NEWALLS (H.J.Z. and J.\'I.), and
INTER/MOBILITY/15/9890527 GREENOX (L.B. and J.\'I.), and by the Nature
Science Foundation of China Grant No. 11574366 (P.C. and
B.G.L.) and the
China Scholarship Council Grant No. 201504910652 (P.C.). 
Additionally, we acknowledge support from the ERC
Consolidator grant MINT with contract number 615759 (M.N.G. and M.B.)
and ARO Grant No. W911NF-16-1-0227 (L.B.). Some figures were prepared
using VESTA\cite{vesta11} and matplotlib,\cite{matplotlib} and we made
extensive use of several crystallographic
servers.\cite{aroyo06a,aroyo06b,aroyo11,campbell06,hatch03} Useful
discussions with Philippe Ghosez and Mael Guennou are gratefully
acknowledged.


%

\end{document}